\documentclass[apj,12pt,preprint]{emulateapj}
\usepackage{lscape}
\shorttitle{Spectropolarimetric Survey of DA White Dwarfs}
\shortauthors{Kawka et al.}
\journalinfo{Accepted for publication in ApJ}
\begin{document}
   \title{Spectropolarimetric survey of hydrogen-rich white dwarf stars}
   \author{A. Kawka\altaffilmark{1,2}, S. Vennes\altaffilmark{3}, G.D. Schmidt\altaffilmark{4}, D.T. Wickramasinghe\altaffilmark{5} and  R. Koch\altaffilmark{2}}
\altaffiltext{1}{Astronomick\'y \'ustav AV \v{C}R, Fri\v{c}ova 298,
 CZ-251 65 Ond\v{r}ejov, Czech Republic}
\altaffiltext{2}{Division of Science and Engineering, 
 Murdoch University, Perth, WA 6150, Australia}
\altaffiltext{3}{Department of Physics and Space Sciences, 150 W. University Blvd, Florida Institute of Technology, Melbourne, FL 32901, USA}
\altaffiltext{4}{Steward Observatory, The University of Arizona, Tucson,
 AZ 85721, USA}
\altaffiltext{5}{Department of Mathematics, Australian National University, Canberra, Australia}

\begin{abstract}
We have conducted a survey of 61 southern white dwarfs searching for magnetic 
fields using Zeeman spectropolarimetry. Our objective is to obtain a magnetic
field distribution for these objects and, in particular, to find white dwarfs
with weak fields. We found one possible candidate (WD 0310$-$688) that may have
a weak magnetic field of $-6.1\pm2.2$ kG. Next, we determine the fraction and 
distribution of magnetic 
white dwarfs in the Solar neighborhood, and investigate the probability of
finding more of these objects based on the current incidence of magnetism
in white dwarfs within 20 pc of the Sun. We have also analyzed the spectra
of the white dwarfs to obtain effective temperatures and surface gravities.
\end{abstract}

\section{Introduction}

White dwarfs are fossil remains of stellar evolution, and a study of
the distribution
of magnetic fields among white dwarfs may help elucidate the role played by
magnetic fields in the evolution of low to intermediate mass stars.

Our current knowledge of the field distribution, i.e., the fraction of
magnetic white dwarfs as a function of field strength (polar, surface
average, or longitudinal field strength), is dictated by a compilation
of surveys of various aims and field sensitivities. Spectroscopic
surveys are useful in probing the Zeeman effect in the Balmer lines of
hydrogen-rich DA white dwarfs,
and may achieve field sensitivity of $\ga 1$MG causing Zeeman splitting of
$\Delta\lambda \ga 20$\AA\ in low-dispersion
spectra \citep{ven1999,fer1998}, or $\gtrsim 10$kG corresponding to
$\Delta\lambda \ga 0.2$\AA\ in high-dispersion
spectra of bright candidates \citep[e.g.,][]{koe1998}. On the other hand, 
low-dispersion spectropolarimetric surveys of DA white dwarfs may achieve
sensitivities of $\gtrsim 10$kG while offering increased
accessibility to fainter white dwarfs \citep{sch1995}.

Similar studies \citep[e.g.,][]{put1997,jor1998,sch2001b} of the
incidence of magnetic fields in non-DA white dwarfs (helium-line DB, or
continuum-like DC) are more difficult because of the weakness
of helium lines relative to hydrogen lines in stars with similar ages,
although the presence of Zeeman-split metal lines may betray 
the presence of a magnetic field in such objects \citep[e.g.,][]{rei2001}.
Dichroism of continuous opacities induces optical circular polarization
of the order of $V/I \approx 0.1B$\%, where $B$ is expressed in MG 
\citep{kem1970,ang1981}. Therefore, polarimetry of DC white dwarfs may achieve 
field sensitivity $\gtrsim 1$MG in high-quality surveys ($V/I\gtrsim 0.1$\%) 
or more generally $\gtrsim 10$MG ($V/I\gtrsim 1$\%).

Spectroscopic and polarimetric surveys have accumulated sufficient 
data to describe the field distribution $\gtrsim1$MG in all nearby white dwarfs 
(DA and non-DA), while only a fraction of the data is available to describe 
the distribution $\gtrsim0.1$MG among brighter hydrogen-rich DA white dwarfs. 
The aim of spectropolarimetric surveys is to complement spectroscopic 
surveys and provide data for fainter DA candidates with even lower fields in 
the 10kG-1MG range. \citet{kaw2003} derived the field distribution in the local 
white dwarf census of \citet{hol2002}. The incidence of magnetic field appears 
constant per decade of field strength from 0.1MG to close to 1000 MG.
Since the estimate of \citet{kaw2003}, which included the low-field magnetic
white dwarf 40 Eri B \citep{fab2000}, \citet{azn2004} added the low-field
white dwarf LTT~9857 (3.1 kG) to the local census (i.e., within 20 pc of the 
Sun) thereby extending the field 
distribution well below 10 kG. The origin of fields in white dwarf stars is 
more difficult to establish quantitatively. \citet{kawthesis} and 
\citet{kaw2004b} revisited some assumptions about the magnetic white dwarf 
space density and the corresponding space density of their likely Ap/Bp 
progenitors. 
They concluded that our current knowledge of stellar formation rate and 
evolutionary time scales cannot account for the present day density of magnetic 
white dwarfs. Low field white dwarfs are also not accounted for, which prompted 
\citet{kaw2004b} to suggest that additional progenitors are required. 

In \S 2 we present spectropolarimetry of 61 white dwarfs with $V \lesssim 16.4$ 
and $\delta < -30^\circ$, among which 55 have $V \lesssim 15.0$ and 5 are in
close binaries. We also present spectropolarimetry of 4 subdwarf B (sdB) stars, 
where two of these were misclassified as white dwarfs. This complements a 
spectropolarimetric survey for magnetic fields among northern hemisphere white 
dwarfs carried out by \citet{sch1995}. That survey sampled some 169 DA white 
dwarfs and resulted in the discovery of four new magnetic white dwarfs with 
fields between $\sim$$10^5$ and $10^9$ G. 
We present our analyses of our observations in \S 3 and
in \S 4 we discuss the properties of the local population of magnetic white
dwarfs. Finally, in \S 5 we summarize our results. Appendix A 
tabulates the properties of all magnetic white dwarfs known to date.

\section{Observations}

\subsection{Spectropolarimetry}

The data for the survey of magnetic fields in southern white dwarfs were
acquired using the 74-inch telescope at the Mt. Stromlo Observatory with the
Steward Observatory CCD Spectropolarimeter. The original instrument is 
described in \citet{sch1992}. The instrument setup was
updated and modified for the 74-inch telescope, with an improved camera
lens and a thinned, back illuminated
LORAL $1200 \times 800$ CCD with near unity quantum efficiency and $< 6$e$^-$
read noise. The $f/18$ Cassegrain telescope beam was adapted to the
$f/9$ spectrograph optics with a small converging lens placed in front
of the slit. A 964 lines mm$^{-1}$ grating blazed at 4639 \AA\ was used which
gave a dispersion of 2.62 \AA\ per pixel.
Circularly polarized spectra of white dwarfs were obtained over a region which
includes H$\alpha$, H$\beta$ and H$\gamma$, with a spectral resolution of
$\sim 9$ \AA. The data were acquired in multiple waveplate sequences.
The length of an exposure, which varied from 360 to 2400 seconds depending on
the brightness of the object and the seeing conditions, is the time required
for one
waveplate sequence to be completed. One waveplate sequence is a series of
four exposures at different quarter-waveplate orientations that produces two
complementary images in opposite senses of circular polarization.
These are used to obtain the degree of circular polarization as a function of 
wavelength $v_\lambda$, as well as the total (unpolarized) spectral flux 
$F_\lambda$. The slit width was generally set at $2.4''$, however it was 
increased to $3.6''$ when the seeing deteriorated.
The observations were conducted on 2000 October 27-29, 
November 3, 5, 25-26, December 1-4, 23-28, 31, 2001 January 1, 18-20, 27,
and February 17, 18, 21, 22 and 25.

\subsection{Complementary Spectroscopy}

Additional spectroscopy of many of the white dwarfs in the survey have been
obtained using the 74-inch telescope with the Cassegrain spectrograph equipped
with a 300 line mm$^{-1}$ grating blazed at 5000 \AA\ and a 2k $\times$ 4k
CCD binned 2 $\times$ 2. This resulted in a wavelength dispersion of
2.85 \AA\ per pixel and a spectral range of about 3500 \AA\ to 6400 \AA\ 
providing a spectral resolution of $\sim 8$ \AA.
The observations were carried out on 1998 May 26, June 18, 2001 February 28, 
March 2, 4, 5, 7, 8, September 13, 15, 16 and October 25 - 28, and 
2002 January 7, 8, March 8 and April 3.
The purpose of obtaining these additional spectra was to have spectral
coverage of the upper Balmer line series, not covered by the spectropolarimeter,
and to constrain the temperature and gravity of the white dwarfs.
For many of the hot white dwarfs in our sample, we have re-analyzed the spectra
from \citet{ven1996}, \citet{ven1997}, \citet{fer1997}, \citet{ven1999} and
\citet{kaw2004}.

\section{Analysis}

The mean longitudinal magnetic field for a specific absorption line is 
measured at each wavelength using the weak-field approximation \citep{ang1973}
\begin{equation}
B_\ell=\frac{v F}{4.67 \times 10^{-13} \lambda^2 \frac{dF}{d\lambda}}
\end{equation}
where $\lambda$ is the wavelength in \AA, $B_\ell$ is the longitudinal magnetic
field strength in G, $F$ is the total spectral flux (i.e., the total intensity, 
$I$) and $v = V/I$ is the degree of circular polarization.
The flux gradient $dF/d\lambda$ is calculated 
from a pseudo-Lorentzian fit to the line profile of the normalized flux.
The fitted line profile is used in this procedure instead of the observed 
profile to reduce the effects of statistical noise in the flux distribution 
and is appropriate when the Zeeman splitting is not resolved.
For a typical white dwarf this is a reasonable
approximation for $B \lesssim 1$ MG.  The quoted $B_\ell$ for a given star and
epoch is computed from the weighted average of measurements at various
wavelength bins across a profile, followed by a weighted average for the lines
observed, here H$\alpha$, H$\beta$, and H$\gamma$.

The uncertainty in $B_\ell$ at each wavelength bin includes two sources of 
noise. The first and most important contribution is derived from the statistical
fluctuation in the circular polarization spectrum. This is measured in the
far wings of the absorption line and converted to an uncertainty in $B_\ell$
(which we call $\sigma_i$) using standard error propagation techniques.
Assuming all errors are statistical, these point-by-point uncertainties
combine to an uncertainty in the value of $B_\ell$ for the entire line as
$\sigma_{noise}=1/\sqrt(\sum 1/\sigma_i^2)$.

The second contribution to the uncertainty of a derived field strength stems 
from the uncertainty in fitting the pseudo-Lorentzian line profile (i.e.,
they are estimated from the root-mean-square deviation of the data from the 
best-fit line profile), and is 
included as an independent noise source $\sigma_{prof}$.  The total 
uncertainty in $B_\ell$ for a given absorption line is then taken to be
$\sigma_{B_\ell}=\sqrt{\sigma^2_{prof} + \sigma^2_{noise}}$.

\begin{deluxetable*}{lcc|lcc}
\tabletypesize{\scriptsize}
\tablecaption{Survey of white dwarfs.\label{tbl_bmeas}}
\tablewidth{0pt}
\tablehead{
\colhead{WD} & \colhead{UT Date} & \colhead{$B_\ell$ (kG)} & \colhead{WD} & \colhead{UT Date} & \colhead{$B_\ell$ (kG)}
}
\startdata
0018$-$339  & 25/11/2000 & $ 0.23\pm 13.52$ & 0859$-$039  & 23/12/2000 & $  -9.41\pm   8.61$ \\
            & 02/12/2000 & $22.86\pm 16.66$ &             & 26/12/2000 & $   5.97\pm   8.00$ \\
            & 04/12/2000 & $-5.21\pm 12.36$ & 0950$-$572  & 04/12/2000 & $   1.25\pm   8.55$ \\
0047$-$524  & 01/12/2000 & $-10.51\pm 13.16$ &             & 24/12/2000 & $   1.03\pm  10.36$ \\
0050$-$332  & 25/12/2000 & $-0.28\pm 9.48$ & 0954$-$710  & 02/12/2000 & $   0.01\pm   4.32$ \\
0106$-$358  & 01/12/2000 & $2.97\pm 17.85$ & 0957$-$666  & 01/12/2000 & $  -9.73\pm  14.42$ \\
            & 24/12/2000 & $-2.15\pm 11.85$ & 0958$-$073  & 27/12/2000 & $   0.01\pm   4.32$ \\
0107$-$342  & 01/12/2000 & $-17.57\pm 12.28$ & 1013$-$050  & 25/02/2001 & $ -18.31\pm  32.80$ \\
            & 24/12/2000 & $-5.17\pm 5.70$ & 1022$-$301\tablenotemark{b} & 21/02/2001 & $  -4.53\pm  27.04$ \\
0126$-$532  & 05/11/2000 & $12.65\pm 17.25$ & 1042$-$690  & 23/12/2000 & $ -20.45\pm   8.96$ \\
            & 26/11/2000 & $7.18\pm 8.08$ & 1053$-$550  & 19/01/2001 & $  -3.74\pm   7.53$ \\
0131$-$164  & 26/12/2000 & $23.27\pm 12.45$ &             & 20/01/2001 & $ -12.19\pm   9.18$ \\
0141$-$675  & 27/10/2000 & $2.93\pm 5.44$ &             & 27/01/2001 & $   1.55\pm   8.13$ \\
            & 03/11/2000 & $-1.68\pm 4.65$ & 1056$-$384  & 01/01/2001 & $   5.90\pm   5.06$ \\
0255$-$705  & 28/10/2000 & $0.27\pm 5.43$ & 1121$-$507  & 17/02/2001 & $  -0.47\pm   7.54$ \\
0310$-$688  & 28/10/2000 & $-6.09\pm 2.24$ &             & 18/02/2001 & $  -6.03\pm   7.98$ \\
0325$-$857  & 23/12/2000 & $7.16\pm 6.27$ & 1153$-$484  & 19/01/2001 & $ -10.41\pm   7.89$ \\
0341$-$459  & 25/11/2000 & $10.73\pm 8.37$ &             & 20/01/2001 & $   2.61\pm   3.86$ \\
            & 26/11/2000 & $4.20\pm 13.62$ & 1236$-$495  & 20/01/2001 & $  -2.58\pm   6.23$ \\
0419$-$487  & 26/11/2000 & $-0.23\pm 9.30$\tablenotemark{a} & 1257$-$723  & 21/02/2001 & $  -7.79\pm   8.01$ \\
            & 24/12/2000 & $5.06\pm 41.61$\tablenotemark{a} & 1323$-$514  & 17/02/2001 & $  -1.84\pm   8.18$ \\
            & 26/12/2000 & $6.08\pm 13.78$\tablenotemark{a} &             & 18/02/2001 & $  -2.24\pm   8.88$ \\
            & 18/01/2001 & $-22.37\pm 27.30$\tablenotemark{a} & 1407$-$475  & 17/02/2001 & $  13.34\pm   8.02$ \\
0446$-$789  & 03/11/2000 & $-7.73\pm 8.40$ & 1425$-$811  & 18/02/2001 & $   2.75\pm   5.13$ \\
            & 24/12/2000 & $-10.60\pm 5.76$ & 1529$-$772\tablenotemark{b} & 25/02/2001 & $  87.05\pm 143.60$ \\
0455$-$282  & 26/12/2000 & $5.57\pm 15.84$ & 1544$-$377  & 18/02/2001 & $  -5.88\pm   7.68$ \\
0501$-$289  & 26/12/2000 & $-5.60\pm 10.35$ &             & 21/02/2001 & $  -0.25\pm   6.45$ \\
0509$-$007  & 25/12/2000 & $6.77\pm 9.78$ & 1616$-$591  & 21/02/2001 & $   7.27\pm   8.92$ \\
0549$+$158  & 26/12/2000 & $11.12\pm 9.18$ &             & 22/02/2001 & $  -8.66\pm   9.96$ \\
0621$-$376  & 02/12/2000 & $-11.12\pm 10.11$ & 1620$-$391  & 21/02/2001 & $  -2.96\pm   2.60$ \\
0646$-$253  & 26/12/2000 & $6.61\pm 6.63$ & 1628$-$873  & 22/02/2001 & $   7.45\pm   6.24$ \\
0652$-$563\tablenotemark{b} & 21/02/2001 & $-4.56\pm 59.48$ & 1659$-$531  & 21/02/2001 & $  -1.06\pm 5.70$ \\
0701$-$587  & 25/11/2000 & $-0.21\pm 7.53$ & 1724$-$359\tablenotemark{b} & 25/02/2001 & $ 11.62\pm 25.06$ \\
            & 24/12/2000 & $-8.44\pm 10.16$ & 2007$-$303  & 27/10/2000 & $   5.50\pm   8.07$ \\
0715$-$703  & 28/12/2000 & $-2.16\pm 17.79$ &             & 29/10/2000 & $   3.68\pm   3.78$ \\
            & 31/12/2000 & $-43.64\pm 21.84$ & 2039$-$682  & 05/11/2000 & $  -6.00\pm 6.40$ \\
            & 01/01/2001 & $-7.06\pm 7.83$ & 2105$-$820  & 28/10/2000 & $  3.39\pm 4.98$ \\
            & 17/02/2001 & $-11.87\pm 10.95$ & 2115$-$560  & 05/11/2000 & $ -1.38\pm 7.05$ \\
0718$-$316  & 02/12/2000 & $51.21\pm 37.70$ & 2159$-$754  & 25/11/2000 & $ -7.75\pm 8.55$ \\
0721$-$276  & 28/12/2000 & $-10.09\pm 12.58$ &             & 02/12/2000 & $ -11.84\pm 7.18$ \\
0732$-$427  & 27/12/2000 & $0.71\pm 7.29$ & 2211$-$495  & 02/12/2000 & $  6.57\pm  5.32$ \\
            & 21/02/2001 & $6.46\pm 7.46$ & 2232$-$575  & 04/12/2000 & $ 11.49\pm 10.17$ \\
0740$-$570  & 27/12/2000 & $-5.48\pm 13.54$ & 2329$-$291  & 24/12/2000 & $  0.19\pm 5.80$ \\
0821$-$252\tablenotemark{b} & 04/12/2000 & $-576.34\pm 106.74$\tablenotemark{a} &  & 25/12/2000 & $ 7.07\pm 6.74$ \\
            & 01/01/2001 & $-698.81\pm 134.97$\tablenotemark{a} & 2331$-$475  & 25/12/2000 & $ -2.18\pm16.84$ \\
            & 25/02/2001 & $-538.19\pm 154.04$\tablenotemark{a} & 2337$-$760  & 03/12/2000 & $ 11.45\pm19.98$ \\
0839$-$327  & 25/11/2000 & $2.89\pm 2.76$ &             & 04/12/2000 & $ 11.19\pm 9.96$ \\
0850$-$617  & 25/11/2000 & $4.58\pm 12.58$ & 2359$-$434  & 27/10/2000 & $  3.41\pm 4.42$ \\
            & 26/11/2000 & $18.53\pm 12.96$ & & & \\
\enddata
\tablenotetext{a}{H$\alpha$ measurement was excluded in the calculation of the mean, see text for details.}
\tablenotetext{b}{EUV-selected ultramassive white dwarfs.}
\end{deluxetable*}

\begin{figure*}
\plotone{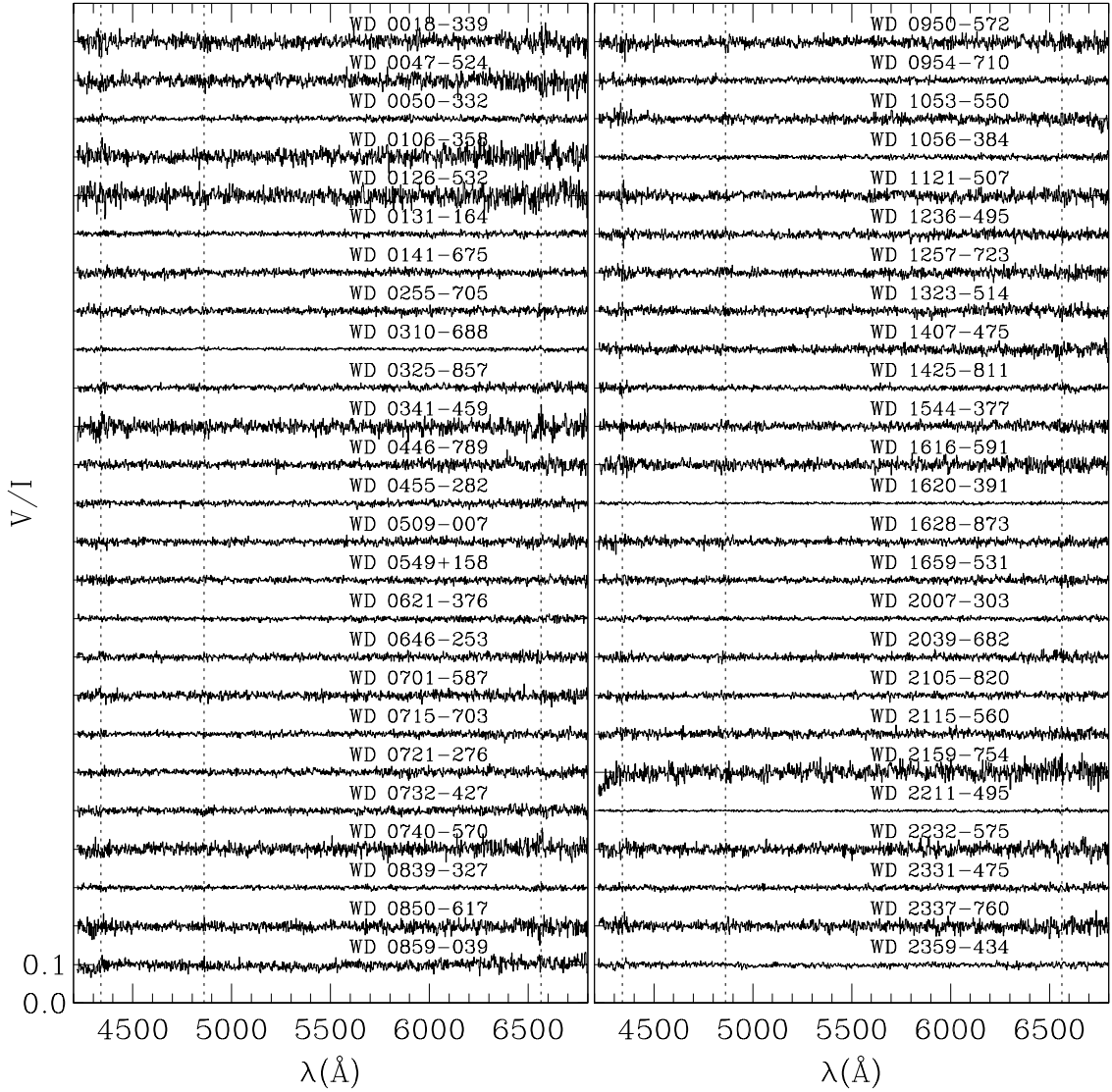}
\caption{Circular polarization spectra of the white dwarfs listed in 
Table~\ref{tbl_survey1}. Each spectrum is separated by 0.10 for better
visualization and the Balmer lines (H$\alpha$, H$\beta$ and H$\gamma$) 
are indicated by the dotted vertical lines.}
\label{fig_specpol}
\end{figure*}

\begin{figure}
\plotone{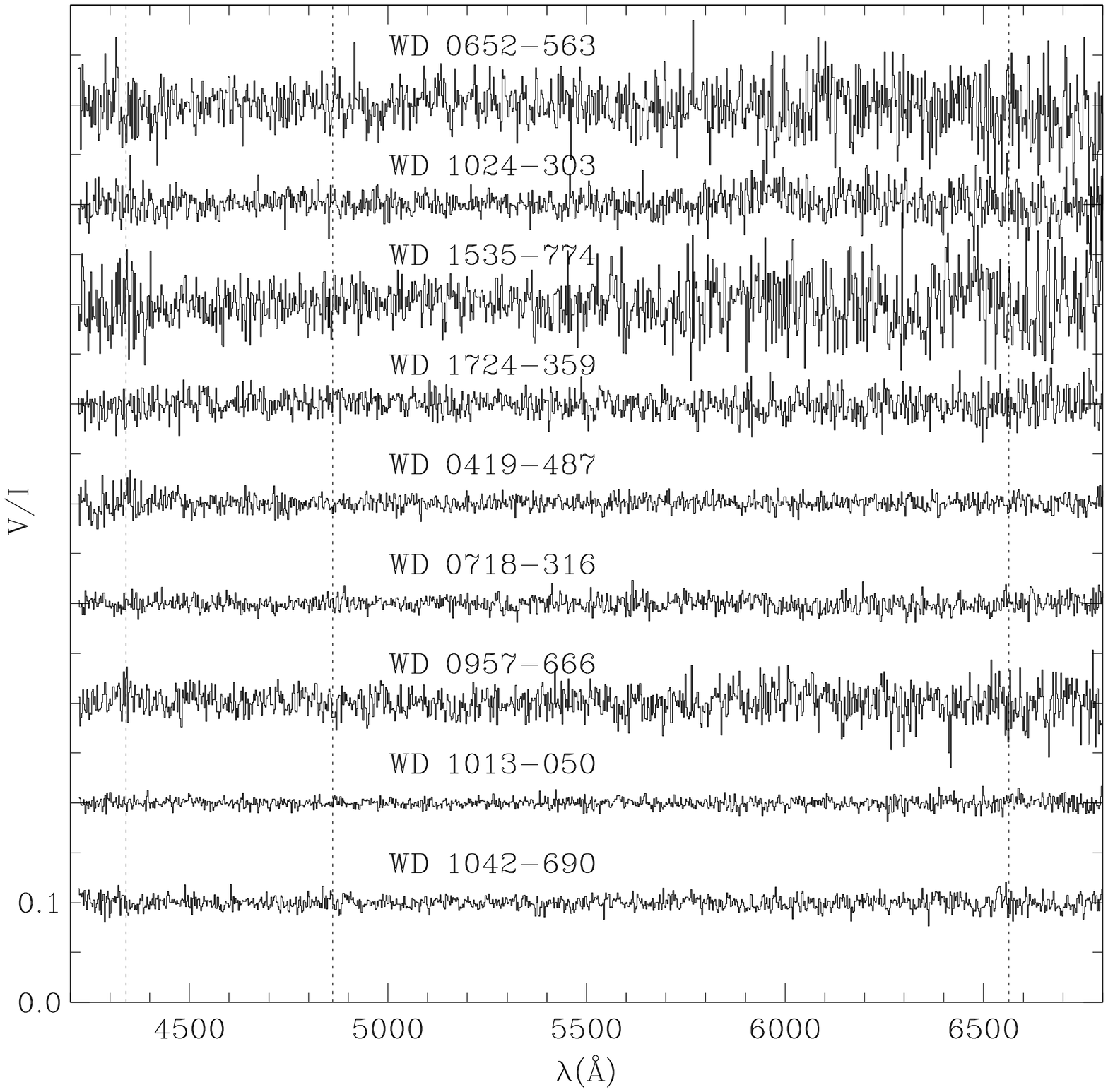}
\caption{Circular polarization spectra of the EUV-selected white dwarfs 
(Table~\ref{tbl_survey1}) and close binary stars 
(Table~\ref{tbl_survey_bin}).
\label{fig_specpol2}}
\end{figure}

\begin{figure}
\includegraphics[width=\columnwidth]{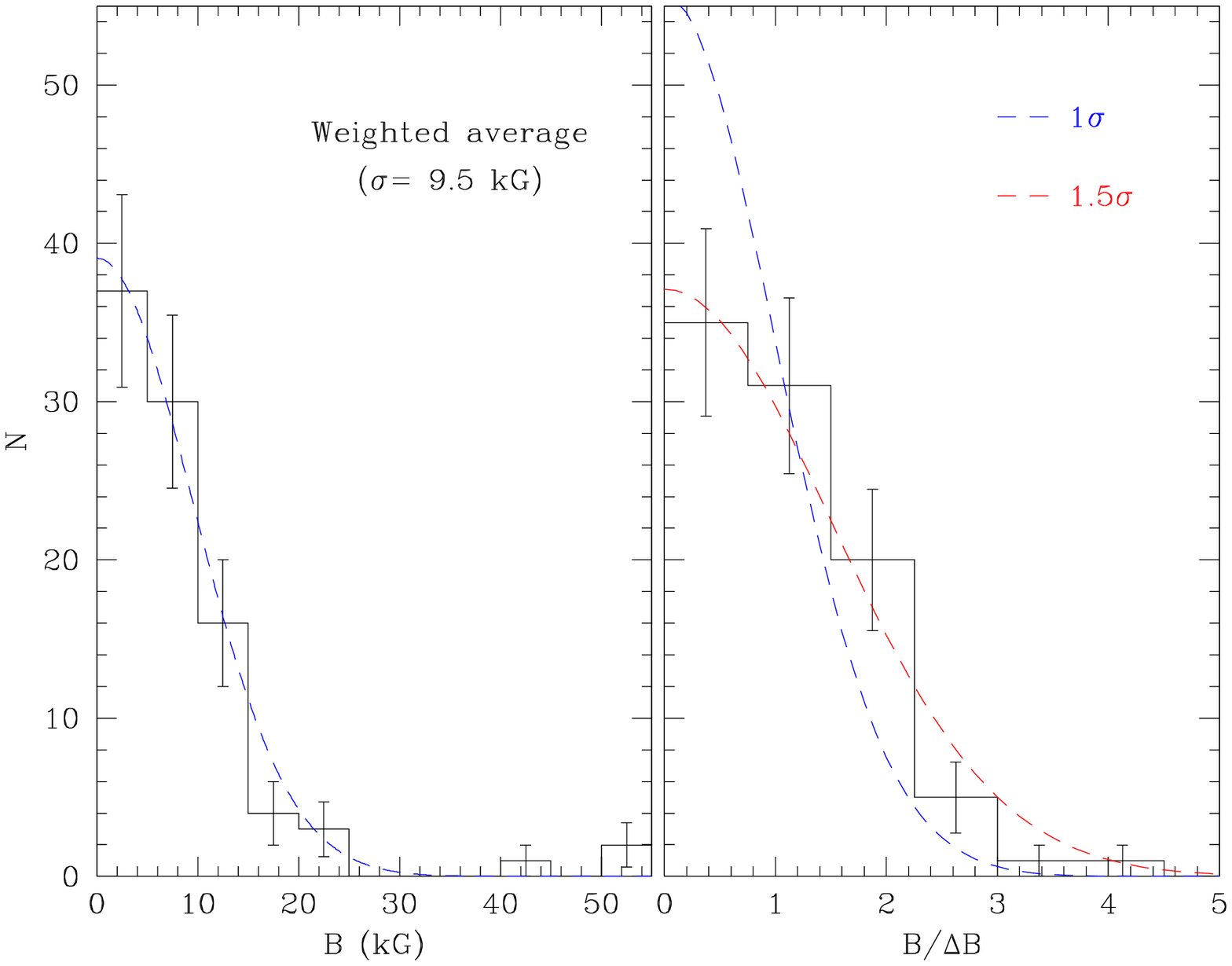}
\caption{Number distribution of measured magnetic field strengths
($N_{\rm total} = 86$) compared to a gaussian
distribution with $\sigma = 9.5$ kG ({\it left}), and the number distribution
of the measurements divided by their error and compared to a normalized gaussian
distribution ({\it right}).
\label{fig_SPOL}}
\end{figure}

The results are tabulated in Table~\ref{tbl_bmeas} and displayed as circular 
polarization spectra in Figures~\ref{fig_specpol} and \ref{fig_specpol2}.
The survey for magnetism yielded no detections of magnetism (those where the
signal-to-noise ratio in the derived field strength clearly exceeds 3.0) with
the possible exception of WD 0310$-$688. For an intrinsically nonmagnetic
sample, one would expect that a histogram of the measured values of $B_\ell$
should approximate a Gaussian whose width is the mean uncertainty 
$\sigma_{B_\ell}$. However, as shown in Figure~\ref{fig_SPOL} the 9.5 kG width
of the distribution is $\sim 50\%$ larger than the mean uncertainty. This
could be due to the stars having weak magnetic fields of the order of a few
kG, but we believe a more likely explanation is that our purely statistical
estimates underrepresent the true uncertainties. For example, seeing and
guiding variations, fluxing errors, and polarimetric calibration uncertainties
can all affect the derived polarization spectra but have not been included in
the analysis above. Similar conclusions were reached for the northern survey by
\citet{sch1995}. Therefore, in this paper all magnetic field measurements have 
been increased by 50\% over their purely statistical values including the values
in Table~\ref{tbl_bmeas}.

\begin{deluxetable*}{llccccc}
\tabletypesize{\scriptsize}
\tablecaption{Survey of white dwarfs.\label{tbl_survey1}}
\tablewidth{0pt}
\tablehead{
\colhead{WD} & \colhead{Other Names} & \colhead{m$_V$} & \colhead{$T_{eff}$} & \colhead{$\log{g}$} & \colhead{M} & \colhead{$d$} \\
\colhead{} & \colhead{} & \colhead{(mag)} & \colhead{(K)} & \colhead{(cgs)} & \colhead{(M$_\odot$)} & \colhead{(pc)} \\
}
\startdata
0018$-$339 & GD 603, BPM 46232             & 14.62 & $19460\pm250$  & $7.87\pm0.05$ & $0.55\pm0.03$ & 63  \\
0047$-$524 & BPM 16274, L219-48            & 14.16 & $17660\pm180$  & $7.82\pm0.04$ & $0.52\pm0.02$ & 49  \\
0050$-$332 & GD 659, EUVE J0053-329        & 13.36 & $35840\pm250$  & $7.90\pm0.05$ & $0.61\pm0.02$ & 60  \\
0106$-$358 & GD 683, EUVE J0108-355        & 15.80 & $29380\pm400$  & $7.86\pm0.10$ & $0.57\pm0.05$ &158  \\
0126$-$532 & BPM 16501, LTT 805            & 14.48 & $16010\pm160$  & $7.94\pm0.04$ & $0.58\pm0.02$ & 48  \\
0131$-$164 & GD 984, EUVE J0134-161        & 13.8  & $45640\pm1220$ & $7.82\pm0.07$ & $0.59\pm0.03$ & 91  \\
0141$-$675 & LTT 934, LHS 145              & 13.90 &  $6460\pm160$  & $8.04\pm0.44$ & $0.61\pm0.24$ &  9  \\
0255$-$705 & LHS 1474, BPM 2819            & 14.08 & $10620\pm90$   & $8.28\pm0.06$ & $0.78\pm0.04$ & 21  \\
0310$-$688 & LB 3303, EG 21                & 11.40 & $15480\pm80$   & $8.02\pm0.02$ & $0.62\pm0.01$ & 11  \\
0325$-$857\tablenotemark{a} & LB 9802      & 13.9  & $15580\pm200$  & $8.36\pm0.05$ & $0.83\pm0.03$ & 27  \\
0341$-$459 & BPM 31594, L300-34            & 15.03 & $11700\pm250$  & $8.20\pm0.10$ & $0.73\pm0.06$ & 40  \\
0446$-$789\tablenotemark{b} & BPM 3523, L31-99 & 13.47& $22760\pm390$  & $7.70\pm0.06$ & $0.48\pm0.03$ & 47  \\
0455$-$282 & MCT 0455-2812                 & 13.95 & $55680\pm1410$ & $7.90\pm0.09$ & $0.65\pm0.04$ &101  \\
0509$-$007 & EUVE J0512-006, RE J0512-004  & 13.83 & $32400\pm320$  & $7.28\pm0.09$ & $0.36\pm0.03$ &107  \\
0549$+$158 & GD 71, LTT 11733              & 13.06 & $33000\pm190$  & $7.84\pm0.04$ & $0.57\pm0.02$ & 51  \\
0621$-$376 & EUVE J0623-376, RE J0623-374  & 12.09 & $59800\pm1810$ & $7.23\pm0.11$ & $0.45\pm0.02$ & 80  \\
0646$-$253 & EUVE J0648-253, RE J0648-252  & 13.65 & $27720\pm310$  & $7.91\pm0.04$ & $0.59\pm0.03$ & 53  \\
0652$-$563\tablenotemark{c} & EUVE J0653$-$564 &16.40&$33480\pm620$  & $8.92\pm0.13$ & $1.16\pm0.06$ &  99  \\
0701$-$587 & BPM 18394, L184-75            & 14.46 & $14800\pm500$  & $8.32\pm0.08$ & $0.81\pm0.05$ & 35  \\
0715$-$703 & EUVE J0715-704, RE J0715-702  & 14.18 & $42640\pm710$  & $7.81\pm0.08$ & $0.58\pm0.03$ &105  \\
0721$-$276 & EUVE J0723-277, RE J0723-274  & 14.52 & $36500\pm400$  & $7.92\pm0.08$ & $0.62\pm0.04$ &102  \\
0732$-$427 & BPM 33039, LTT 2884           & 14.16 & $14540\pm370$  & $8.15\pm0.06$ & $0.70\pm0.04$ & 33  \\
0740$-$570 & BPM 18615, L185-53            & 15.06 & $19760\pm290$  & $8.20\pm0.06$ & $0.74\pm0.04$ & 62  \\
0800$-$533 & BPM 18764, L242-83            & 15.76 & $20010\pm500$ & $7.86\pm0.08$ & $0.55\pm0.04$ &110 \\
0821$-$252\tablenotemark{c,d} & EUVE J0823$-$254 & 16.40 & $39800\pm1400$ & $9.23\pm0.17$ & $1.27\pm0.06$ &  81  \\
0839$-$327\tablenotemark{e} & LFT 600, LTT 3218 & 11.90 &  $9340\pm60$  & $8.11\pm0.06$ & $0.66\pm0.04$ & 7  \\
0848$-$730 & BPM 5102, L63-60              & 15.30 & $16800\pm350$  & $7.96\pm0.08$ & $0.59\pm0.04$ & 72  \\
0850$-$617 & BPM 5109, L139-26             & 14.73 & $19620\pm340$  & $8.04\pm0.06$ & $0.64\pm0.03$ & 60  \\
0859$-$039 & EUVE J0902-041, RE J0902-040  & 13.19 & $23560\pm290$  & $7.89\pm0.04$ & $0.57\pm0.02$ & 38  \\
0950$-$572 & BPM 19738, L189-36            & 14.94 & $13900\pm600$  & $7.84\pm0.10$ & $0.52\pm0.05$ & 57  \\
0954$-$710 & BPM 6082, L64-27              & 13.48 & $14280\pm240$  & $7.71\pm0.06$ & $0.46\pm0.03$ & 32  \\
1022$-$301\tablenotemark{c} & EUVE J1024$-$303, RE J1024$-$302 & 16.09 & $34700\pm700$  & $9.05\pm0.12$ & $1.21\pm0.05$ &  78  \\
1053$-$550 & LTT 4013, BPM 20383           & 14.32 & $14000\pm330$  & $8.11\pm0.10$ & $0.67\pm0.06$ & 36  \\
1056$-$384 & EUVE J1058-387, RE J1058-384  & 14.08 & $27800\pm420$  & $8.02\pm0.08$ & $0.65\pm0.04$ & 60  \\
1121$-$507 & BPM 20912, L251-24            & 14.86 & $15620\pm220$  & $7.93\pm0.06$ & $0.57\pm0.03$ & 57  \\
1223$-$659 & BPM 7543, L104-2              & 13.97 & $7740\pm70$ & $8.13\pm0.11$ & $0.67\pm0.07$ & 13  \\
1236$-$495 & LTT 4816, LFT 931             & 13.96 & $11870\pm130$  & $8.84\pm0.04$ & $1.11\pm0.02$ & 15  \\
1257$-$723 & BPM 7961, L69-47              & 15.18 & $16380\pm430$  & $7.98\pm0.10$ & $0.60\pm0.05$ & 66  \\
1323$-$514 & LFT 1004, LTT 5178            & 14.60 & $18280\pm220$  & $7.92\pm0.04$ & $0.57\pm0.02$ & 58  \\
1407$-$475 & BPM 38165, L332-123           & 14.31 & $21080\pm400$  & $8.00\pm0.06$ & $0.63\pm0.03$ & 54  \\
1425$-$811 & BPM 784, LTT 5712             & 13.75 & $12600\pm240$  & $8.17\pm0.06$ & $0.71\pm0.04$ & 24  \\
1529$-$772\tablenotemark{c} & EUVE J1535$-$774 & 16.40 & $49200\pm1800$ & $8.95\pm0.16$ & $1.18\pm0.07$ & 119  \\
1544$-$377 & LTT 6302, L481-60             & 12.80 & $10840\pm80$   & $8.27\pm0.06$ & $0.77\pm0.04$ & 12  \\
1616$-$591 & BPM 24047, LTT 6501           & 15.08 & $13950\pm260$  & $7.81\pm0.07$ & $0.51\pm0.03$ & 62  \\
1620$-$391 & CD-38 10980, EUVE J1623-392   & 11.01 & $23860\pm390$  & $8.02\pm0.06$ & $0.64\pm0.03$ & 13  \\
1628$-$873 & BPM 890, L8-61                & 14.58 & $11160\pm190$  & $8.29\pm0.07$ & $0.78\pm0.04$ & 29  \\
1659$-$531 & BPM 24601, L268-92            & 13.47 & $14660\pm190$  & $8.02\pm0.04$ & $0.62\pm0.02$ & 27  \\
1709$-$575 & LTT 6859, BPM 24723           & 15.10 & $16800\pm320$ & $7.87\pm0.07$ & $0.54\pm0.03$ & 70  \\
1724$-$359\tablenotemark{c} & EUVE J1727$-$360, RE J1727$-$355 & 15.46 & $32000\pm250$  & $9.00\pm0.06$ & $1.20\pm0.03$ &  58  \\
1953$-$715 & LTT 7875, BPM 12843           & 15.15 & $18200\pm480$ & $7.97\pm0.10$ & $0.60\pm0.05$ & 72  \\
2007$-$303 & LTT 7987, L565-18             & 12.18 & $14840\pm170$  & $7.92\pm0.04$ & $0.57\pm0.02$ & 16  \\
2039$-$682 & LTT 8190, BPM 13491           & 13.53 & $16190\pm220$  & $8.47\pm0.04$ & $0.90\pm0.03$ & 22  \\
2105$-$820 & BPM 1266, LTT 8381            & 13.62 & $10660\pm90$   & $8.26\pm0.06$ & $0.76\pm0.04$ & 18  \\
2115$-$560 & LTT 8452, BPM 27273           & 14.28 & $9740\pm80$    & $8.30\pm0.08$ & $0.79\pm0.05$ & 20  \\
2159$-$754 & LTT 8816, BPM 14525           & 15.06 & $9040\pm80$    & $8.95\pm0.12$ & $1.17\pm0.04$ & 14  \\
2211$-$495 & EUVE J2214-493, RE J2214-491  & 11.71 & $63840\pm1550$ & $7.45\pm0.10$ & $0.51\pm0.02$ & 57  \\
2232$-$575 & LTT 9082, BPM 27891           & 14.96 & $15750\pm290$  & $7.75\pm0.08$ & $0.48\pm0.04$ & 67  \\
2331$-$475 & EUVE J2334-472, RE J2334-471  & 13.42 & $52060\pm1670$ & $7.92\pm0.12$ & $0.65\pm0.05$ & 75  \\
2336$-$079 & GD 1212, GJ 4355              & 13.75 & $11010\pm210$  & $8.05\pm0.15$ & $0.63\pm0.09$ & 23  \\
2337$-$760 & LTT 9648, BPM 15727           & 14.66 & $13800\pm280$  & $7.41\pm0.08$ & $0.32\pm0.04$ & 63  \\
2351$-$368 & LHS 4041, LTT 9774            & 15.1  & $14540\pm320$ & $8.00\pm0.06$ & $0.61\pm0.03$ & 57  \\
2359$-$434\tablenotemark{b} & LTT 9857, BPM 45338 & 13.05 & $8570\pm50$ & $8.60\pm0.06$ & $0.98\pm0.04$ & 7 \\
\enddata
\tablenotetext{a}{Companion to the magnetic white dwarf EUVE J0317-855.}
\tablenotetext{b}{Weak magnetic field \citep{azn2004}.}
\tablenotetext{c}{EUV-selected ultramassive white dwarfs.}
\tablenotetext{d}{Magnetic.}
\tablenotetext{e}{Suspected double degenerate.}
\end{deluxetable*}

\begin{deluxetable*}{llrcrrcc}
\tabletypesize{\scriptsize}
\tablecaption{Close binary stars.\label{tbl_survey_bin}}
\tablewidth{0pt}
\tablehead{
\colhead{WD} & \colhead{Other Names} & \colhead{m$_V$} & \colhead{Spectral Type} & \colhead{$T_{eff}$} & \colhead{$\log{g}$} & \colhead{References} \\
\colhead{} & \colhead{} & \colhead{(mag)} & \colhead{} & \colhead{(K)} & \colhead{(cgs)} & \colhead{} \\
}
\startdata
0419-487 & LTT 1951, BPM 31852          & 14.36 & DA+dMe  &  $7005\pm140$   & $7.72\pm0.19$ & 1 \\
0718-316 & EUVE J0720-317, RE J0720-318 & 14.82 & DAO+dMe &  $52400\pm1800$ & $7.68\pm0.10$ & 2 \\
0957-666 & BPM 6114, L101-26            & 14.60 & DD      &  $27047\pm398$  & $7.28\pm0.08$ & 1 \\
1013-050 & EUVE J1016-053               & 14.20 & DAO+dMe &  $54800\pm1300$ & $7.70\pm0.10$ & 2 \\
1042-690 & BPM 6502, LTT 3943           & 13.09 & DA+dMe  &  $19960\pm400$  & $7.86\pm0.09$ & 3 \\
\enddata
\tablerefs{(1) \citet{bra1995}; (2) \citet{ven1997}; (3) This work}
\end{deluxetable*}

Balmer line spectra provide insights into the temperature and density 
structure of white dwarf atmospheres represented by the parameters  
$T_{\rm eff}$ and $\log{g}$. We computed a new grid of models supporting a 
($T_{\rm eff}$, $\log{g}$) analysis: the grid of models extend from
$T_{\rm eff} = 4500$ to 6500 K (in steps of 500 K) at $\log{g} = 7.0$, 8.0, and
9.0, from $T_{\rm eff} = 7000$ to 16000 K (in steps of 1000 K), from 18000 to
32000 K (in steps of 2000 K), and from 36000 to 84000 K (in steps of 4000 K) at
$\log{g} = 7.0$ to 9.5 (in steps of 0.25 dex). 
Convective energy transport in cooler atmospheres is included by applying
the Schwarzschild stability criterion and by using the mixing-length formalism 
described by \citet{mih1978}, where we have assumed the ML2 parameterization of
the convective flux \citep{fon1981} and adopting $\alpha = 0.6$ 
\citep{ber1992b}. The equation of convective energy transport was 
fully linearized within the Feautrier solution scheme and subjected to the
constraint that $\mathcal{F}_{\rm total} = \sigma_R T^4_{\rm eff} = 
\mathcal{F}_{\rm conv}+\mathcal{F}_{\rm rad}$.
The dissolution of the hydrogen energy levels in the high-density atmospheres
of white dwarfs was calculated using the formalism of  \citet{hum1988}
and following the treatment of \citet{hub1994}. See \citet{kaw2006} for more
detail of the procedure\footnote{Note that equation (2) in \citet{kaw2006} should be $\beta = K_n\Big(\frac{Z^3}{16n^4}\Big)\Big(\frac{4\pi a_0^3}{3}\Big)^{-\frac{2}{3}}N_e^{-1}N_{\rm ion}^{\frac{1}{3}}$}. The calculated level occupation
probabilities are then explicitly included in the calculation of the line and 
continuum opacities. The Balmer line profiles are calculated using the tables 
of Stark-broadened \ion{H}{1} line profiles of \citet{lem1997} convolved with 
normalized resonance line profiles.

The observed Balmer lines (H$\alpha$/H$\beta$ to H$\epsilon$/H8) were fitted 
with model spectra using $\chi^2$ minimization techniques and the quoted 
uncertainties are statistical only ($1\sigma$) and do not take into account 
possible systematic effects in model calculations or data acquisition and 
reduction procedures. We used the mass-radius relations of \citet{ben1999} 
with a hydrogen envelope of $M_H/M_* = 10^{-4}$ and a metallicity of Z=0 to 
convert the ($T_{\rm eff}$, $\log{g}$) measurements into white dwarf masses.
For two white dwarfs (WD 0621-376 and WD 2211-495) we have used the mass-radius 
relations of \citet{woo1995} to determine their masses, since the mass-radius 
relations of \citet{ben1999} did not extend to the very high temperatures of 
these stars.

\begin{deluxetable*}{llrcrccc}
\tabletypesize{\scriptsize}
\tablecaption{Stars from the survey that are not DA white dwarfs.\label{tbl_survey_oth}}
\tablewidth{0pt}
\tablehead{
\colhead{WD} & \colhead{Other Names} & \colhead{m$_V$} & \colhead{Spectral Type} & \colhead{$T_{eff}$} & \colhead{$\log{g}$} & $\log{(N_{He}/N_H)}$ & \colhead{References} \\
\colhead{} & \colhead{} & \colhead{(mag)} & \colhead{} & \colhead{(K)} & \colhead{(cgs)} & \colhead{} & \colhead{} \\
}
\startdata
0107-342 & GD 687, MCT 0107-3416         & 13.93 & sdB & $24350\pm 100$ & $5.32\pm0.02$ & -2.38 & 1 \\
0501-289 & MCT 0501-2858, EUVE J0503-288 & 13.9  &  DO & $68600\pm1800$ & $7.20\pm0.07$ & $\ga 1.7$ & 2 \\
0958-073 & PG 0958-073, GD 108           & 13.56 & sdB & $27760\pm670$  & $5.60\pm0.11$ & $\la -3.0$ & 3 \\
1153-484 & BPM 36430, L 325-214          & 12.86 & sdB & $30080\pm660$  & $5.15\pm0.10$ & $\la -3.0$ & 3 \\
2329-291 & GD 1669                       & 13.88 & sdB & $34126\pm 100$ & $5.77\pm0.02$ & -1.36 & 1 \\
\enddata
\tablerefs{(1) \citet{lis2005}; (2) \citet{ven1998}; (3) This work}
\end{deluxetable*}

The white dwarfs from the present study are presented in 3 different tables.
Table~\ref{tbl_survey1} presents the DA white dwarfs that were assumed to 
follow single star evolution, white dwarfs in wide binaries are also included 
in this table. This table includes EUVE-selected ultramassive white dwarfs, 
which are indicated. White dwarfs in close binaries are presented in 
Table~\ref{tbl_survey_bin}. And finally Table~\ref{tbl_survey_oth} presents 
the non-DA stars in our spectropolarimetric survey, which comprise of one DO 
white dwarf and 4 sdB stars.

\subsection{Single white dwarfs}

Table~\ref{tbl_survey1} presents the observed white dwarfs with their
apparent magnitude, effective temperature, surface gravity and the
distance. The distance was calculated from the apparent and absolute magnitudes.
For 55 of these stars, spectropolarimetry was obtained.
We have used the complementary spectra that cover the higher Balmer lines to
determine their effective temperature and surface gravity by comparing these
spectra to a grid of synthetic model spectra. Figure~\ref{fig_balmer} shows
the Balmer line profile fits to the observed spectra, which have not been
previously published.

\begin{figure*}
\plotone{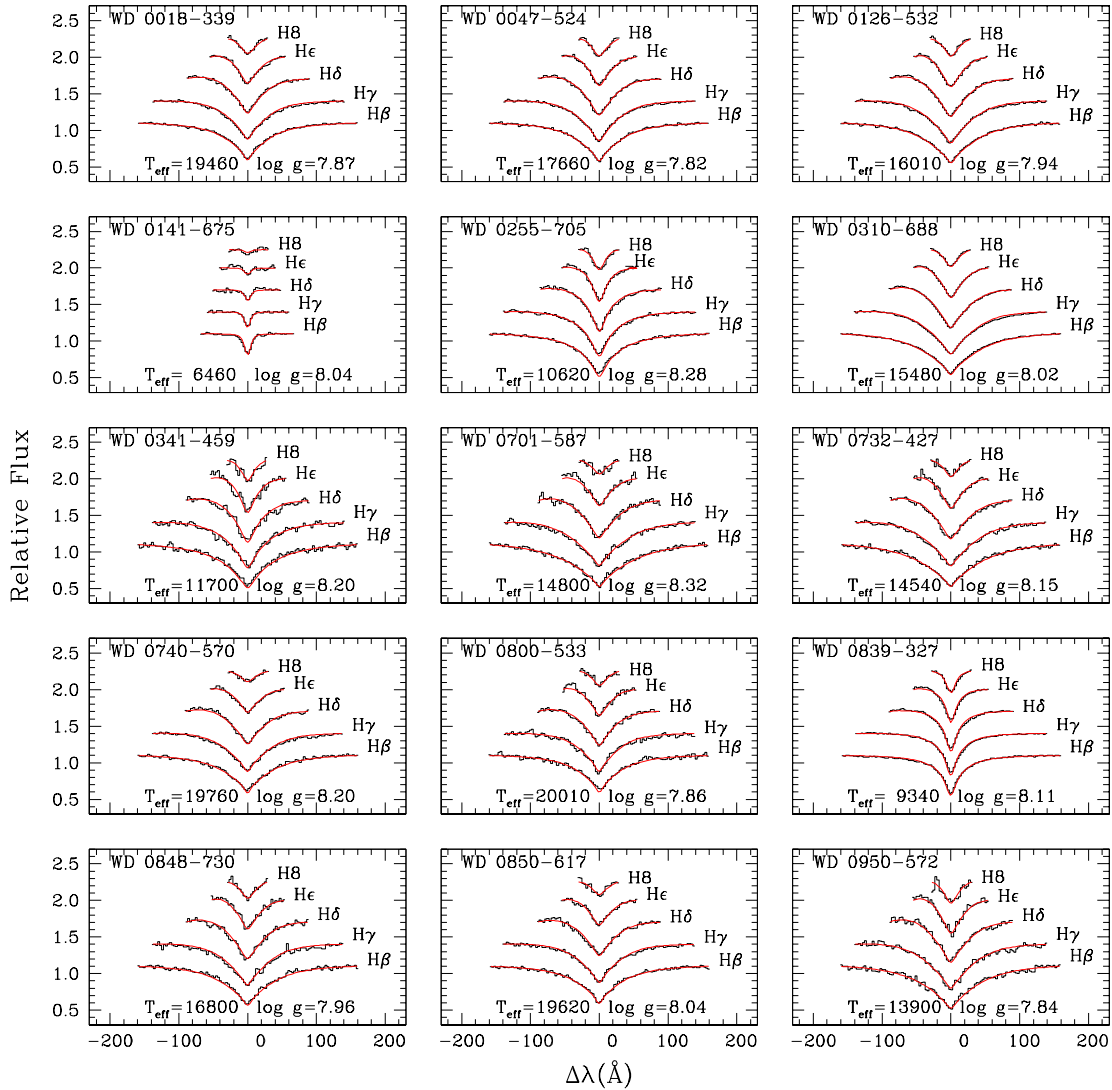}
\caption{Balmer line profile fits to the white dwarfs listed in 
Table~\ref{tbl_survey1}.}
\label{fig_balmer}
\end{figure*}

\addtocounter{figure}{-1}
\begin{figure*}
\plotone{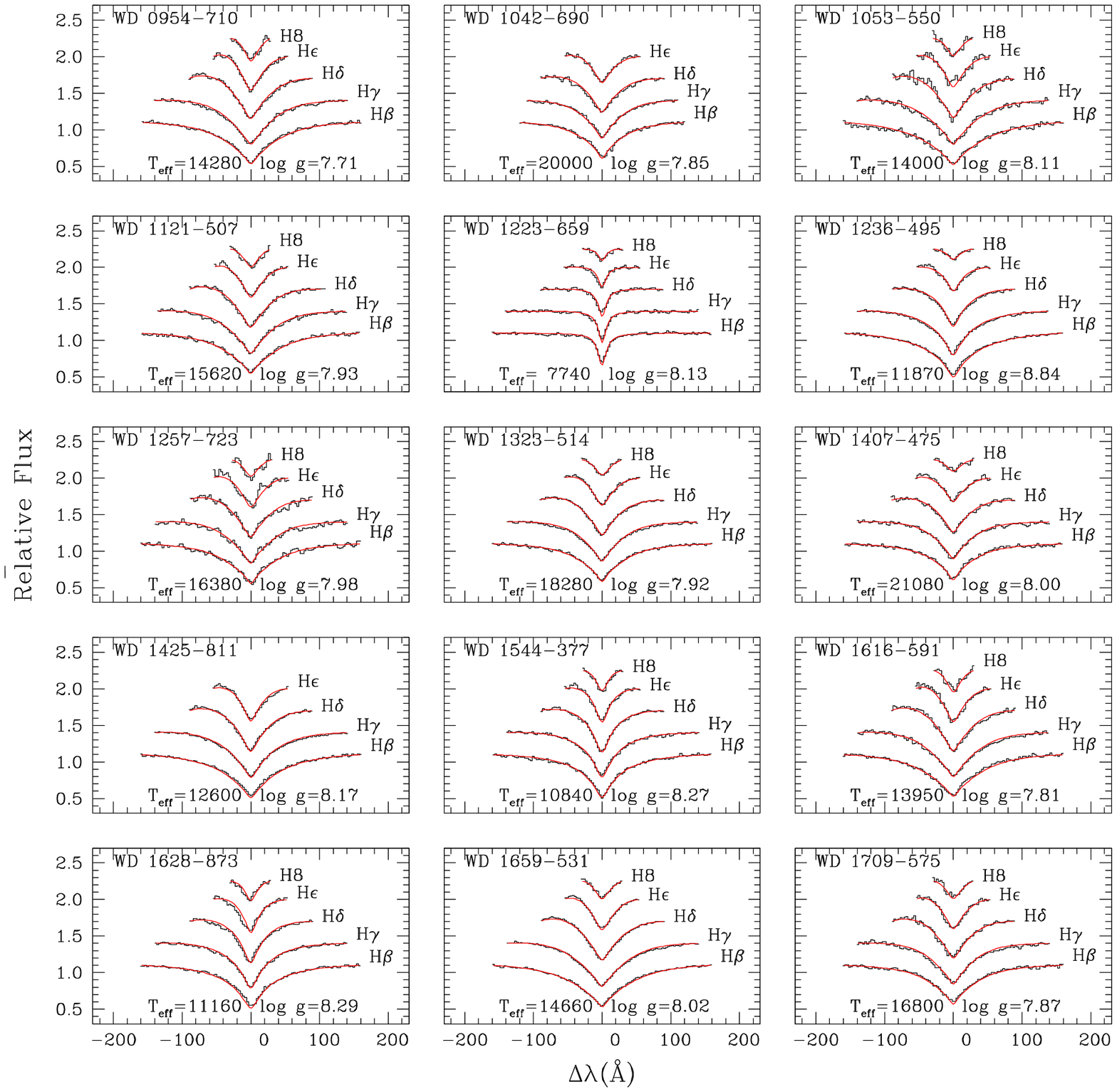}
\caption{Continued}
\end{figure*}

\addtocounter{figure}{-1}
\begin{figure*}
\plotone{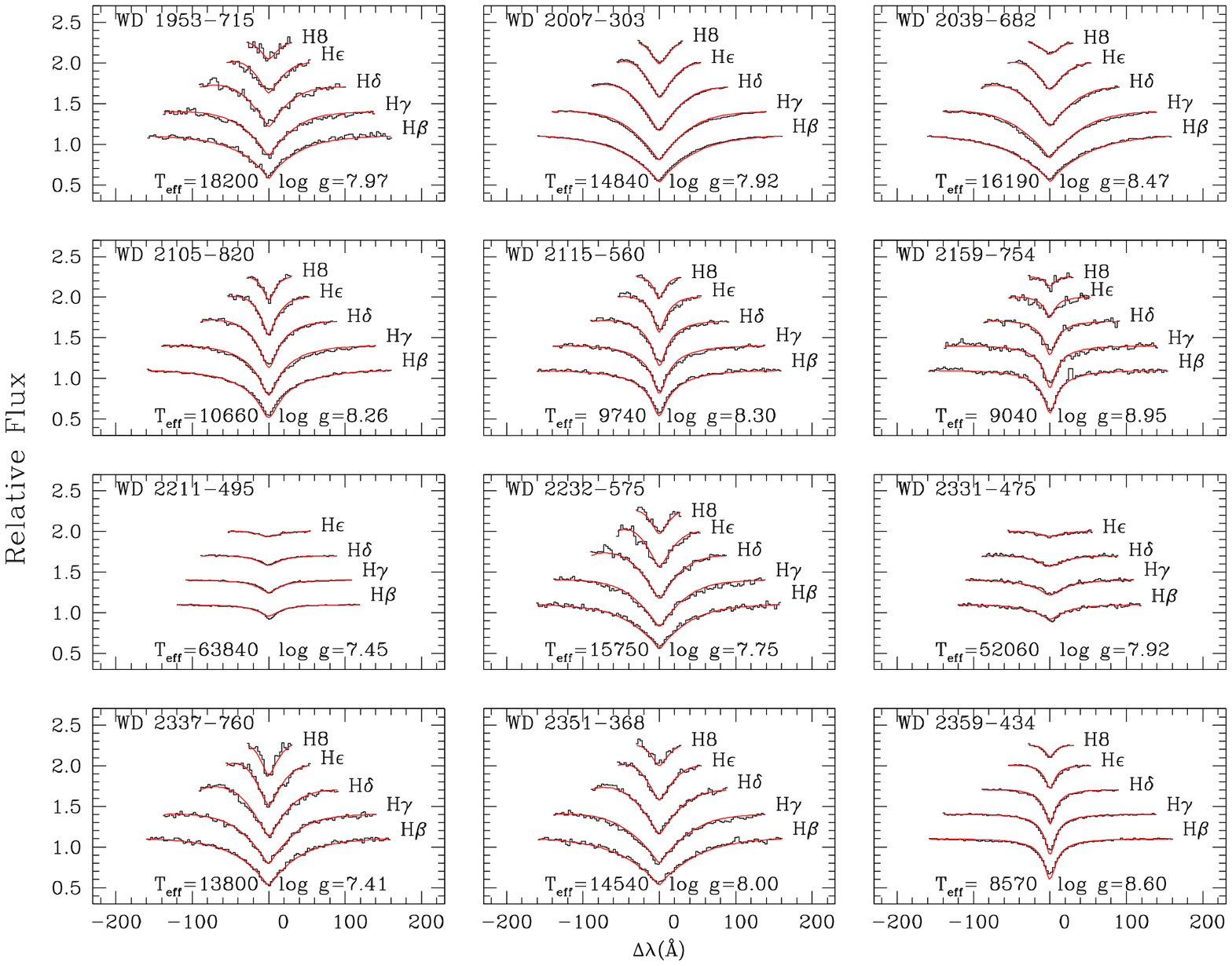}
\vspace{-3cm}
\caption{Continued}
\end{figure*}

Magnetic properties of several stars in this spectropolarimetric survey have
been discussed in the literature, therefore these properties will be summarized
and compared to the results of this study.

\begin{figure}
\plotone{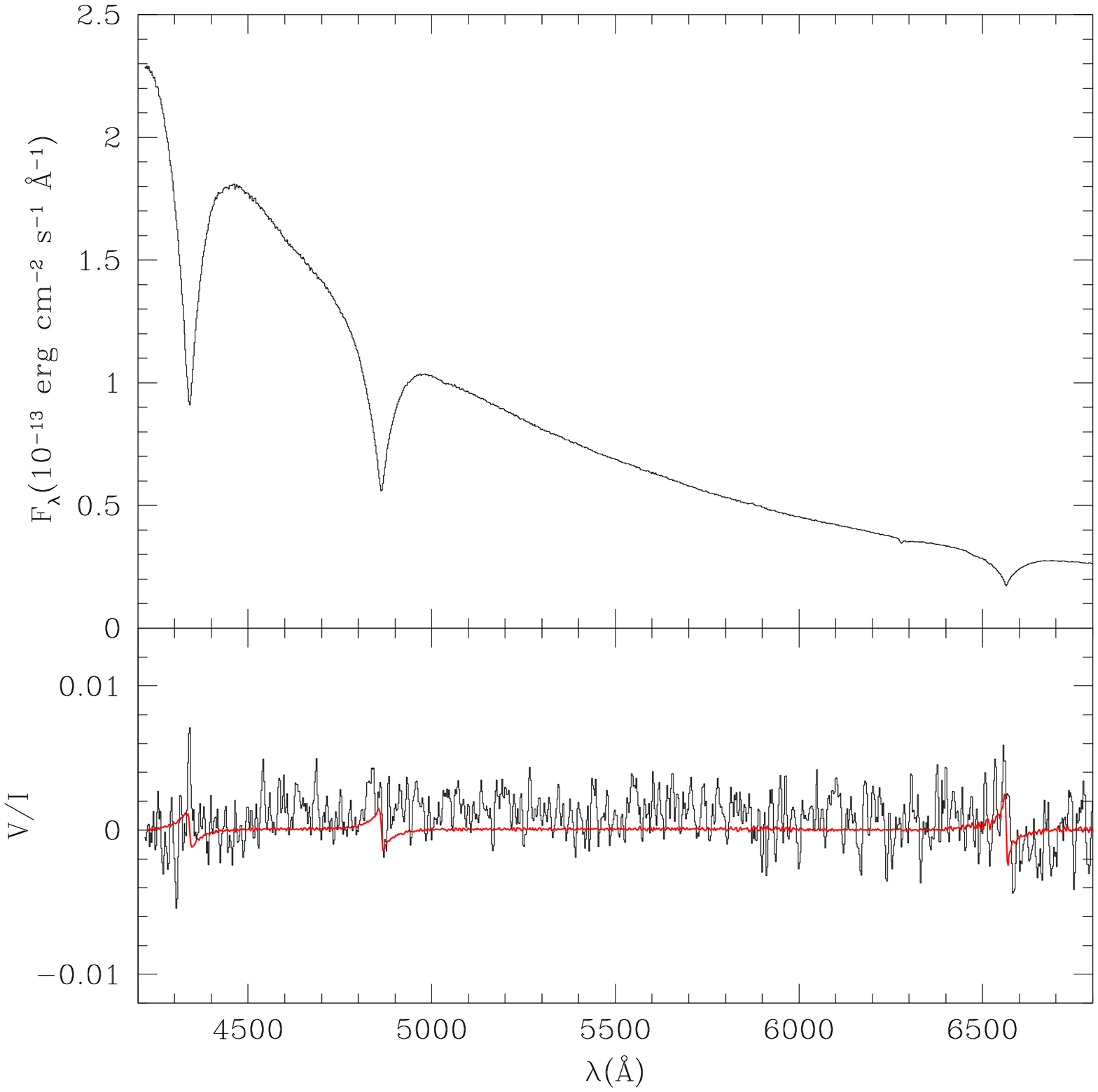}
\caption{The flux ({\it top}) and circular polarization ({\it bottom}) spectra
of WD 0310$-$688. The polarization spectrum is compared to a calculated
circular polarization spectrum assuming a longitudinal field of -6.1 kG.
\label{fig_spol_EG21}}
\end{figure}

\paragraph{WD 0310-688:} The circular polarization shows a hint of the 
presence of a weak magnetic field, but the measurement of $-6.1\pm2.2$ kG
does not exceed $3\sigma$ significance. Figure~\ref{fig_spol_EG21} shows the 
flux and circular polarization spectra of WD~0310$-$688. Note that in a 
gaussian distribution 0.26\% of the objects lie outside $3\sigma$, 
and therefore the probability that we will have at least one measurement above 
the $3\sigma$ limit in our sample of 61 white dwarfs is 14.7\%. We have checked 
the individual measurements for
each line and we found that the magnetic field at H$\gamma$ was significantly 
lower and of opposite sign to the measurements obtained at H$\alpha$ and 
H$\beta$. However, at H$\gamma$ the Zeeman effect is much weaker and the 
signal-to-noise ratio in this region is much lower. If we only consider the
measurements at H$\alpha$ and H$\beta$, then the field measurement would be
$-7.3\pm2.4$, and still barely a $3\sigma$ detection. \citet{azn2004} 
also obtained spectropolarimetry for this star and did not detect the presence 
of a magnetic field down to a limit of 0.5 kG. Different orientations of the 
magnetic field as a result of stellar rotation can produce varying field
strengths \citep[e.g. WD~0009+501,][]{val2005}, therefore further
spectropolarimetric observations are required to confirm the presence of a
magnetic field in WD 0310$-$688.

\paragraph{LB 9802:} This white dwarf is the visual companion to the high-field
ultramassive white dwarf, EUVE J0317$-$85.5. We find that the longitudinal field
measurement of LB 9802 is $7.2\pm6.3$ kG, implying that this is a
non-magnetic white dwarf. Figure~\ref{fig_specpol} shows the flux and 
polarization spectra of LB 9802. The magnetic companion EUVE J0317$-$85.5 was 
found to vary over a 12 minute ($725.4\pm0.9$ seconds) cycle \citep{bar1995}. 
Using far-ultraviolet spectroscopy \citet{bur1999} found that the magnetic
field of EUVE J0317$-$85.5 (RE J0317$-$853) varied between 180 and 800 MG over
the surface of the white dwarf assuming a multipolar expansion of the field.
Using spectropolarimetry and {\it EUVE} photometry \citep{ven2003} were able to
improve the period to $725.727\pm0.001$ seconds. Vennes et al. 
also suggest that EUVE J0317$-$85.5 has an underlying surface magnetic field 
of $\sim 185$ MG with a magnetic spot with a surface field strength of 
$\sim 425$ MG. These two stars have a projected separation of about 200 AU, 
implying a common origin. However, the more massive star is much hotter than 
the less massive and cooler LB 9802, which suggests an age disparity between 
the two stars. A possible explanation for this age disparity is that
EUVE J0317$-$85.5 is a result of a double degenerate merger.

\paragraph{WD 0446-789:} This star has recently been observed by
\citet{azn2004}. Using spectropolarimetry they found this white dwarf to have a
longitudinal magnetic field of $4.283\pm0.640$ kG. The detection of a low
strength magnetic field in a white dwarf means that a population of low
magnetic field white dwarfs may exist. Such a population would contribute
toward a wider $B/\Delta B$ distribution shown in Figure~\ref{fig_SPOL}. Our 
survey was not sensitive enough to detect fields of a few kG.

\paragraph{WD 0621-376:} \citet{ven1999} limited the magnetic field to 30 kG
from the narrow H$\alpha$ core. We measured a longitudinal field of
$-11.1\pm10.1$ kG. Assuming that $B_p = B_{\ell}/(0.4\cos{i})$ and
that $B_p = 30$ kG, then the longitudinal magnetic field can be at most
12 kG, which clearly lies within $2\sigma$ of our measurement.

\paragraph{WD 0839-327:} This object is a possible double degenerate system
\citep{bra1990}. These authors did not find significant radial velocity
changes, however they found significant line profile variations. Using
the trigonometric parallax ($\pi = 112.7\pm9.7$ mas) from the Yale 
Trigonometric Catalog implies a 
distance of $8.87\pm0.77$ pc, which is slightly further than the distance
obtained from our spectroscopic fit (i.e., 7 pc). Assuming this is a double 
degenerate,
we will assume that the determined parameters (i.e., $T_{\rm eff} = 9340$ K 
and $\log{g}=8.11$) are for the primary (i.e., the brighter component).
To approximate the temperature of the secondary component, we first calculate
the absolute magnitude of the primary from the temperature and gravity 
determined from the spectroscopic fit. Therefore, the absolute magnitude of the
primary is $M_V = 12.58$. Next, the total absolute magnitude for the system
calculated from the distance ($d = 8.9$ pc) and apparent magnitude ($V=11.90$ 
mag) is $M_V = 12.15$. The difference in $M_V$ suggests that the secondary 
component would be required to have $M_V = 13.4$ and assuming $\log{g}=8$
this would correspond to an effective temperature of $\sim 7500$ K.
 
\paragraph{WD 0859-039:} This star has also been observed
by \citet{azn2004}.
Their spectropolarimetric observations did not reveal the presence of a
magnetic field in this star down to a limit of $\sim 0.5$ kG,
which supports our measurements and non-detection.

\paragraph{WD 1544-377:} This star is the common proper-motion companion to 
the bright G6V star, HD 140901. \citet{koe1998} limited the magnetic field to 
20 kG from the narrow H$\alpha$ core. We can therefore expect the longitudinal
field to be at most 8 kG if we assume a simple dipole for the magnetic field.
Both our longitudinal field measurements of $-5.9\pm7.7$ kG and $-0.25\pm6.5$ kG
lie within 8 kG.

\paragraph{WD 1620-391:} \citet{koe1998} limited the magnetic field to 10 kG
from the narrow H$\alpha$ core, therefore assuming a simple dipole the 
longitudinal field can be at most 4 kG. We measured a longitudinal field of
$-3.0\pm2.6$ kG and within $2\sigma$ of this measurement we cannot 
place any tight constraints on the inclination.

\paragraph{WD 1659-531:} This star is the common-proper motion companion to the
bright F star, HD 153580. \citet{koe1988} limited the magnetic field to 25 kG
from the narrow H$\alpha$ core, and assuming a simple dipole the longitudinal
field can be at most 10 kG. We measured a longitudinal field of
$-1.1\pm5.7$ kG and within $2\sigma$ of this measurement we cannot 
place any tight constraints on the inclination.

\paragraph{WD 2007-303:} \citet{koe1998} limited the magnetic field to 10 kG
from the narrow H$\alpha$ core and assuming a simple dipole the longitudinal
field can be at most 4 kG. We have obtained two measurements of the longitudinal
field at different epochs ($5.5\pm8.1$ kG and $3.7\pm3.8$ kG) which cannot be
used to place any tight constraints on the inclination.

\paragraph{WD 2039-682:} A broadened H$\alpha$ core was observed by
\citet{koe1998}. They fitted a rotationally broadened profile of 
$v \sin{i} = 80$ km s$^{-1}$
to the core, but they also suggested that a magnetic field of $\approx 50$ kG
could cause the broadening.
Our longitudinal field measurement
of $-6.0\pm6.4$ kG suggests the broadening is most likely due to rotation,
however a magnetic field (i.e., $B_\ell < 12.8$ kG $= 2\sigma$) may still be 
present if it is viewed at high inclination ($i>50^\circ$).
Also a magnetic spot on the surface of the white dwarf may exist but which
was hidden from view when this object was observed during the survey.
For example, WD~1953-011 is a known magnetic white dwarf which appears to have 
a magnetic spot on its surface \citep{max2000}.

\paragraph{WD 2105-820:} \citet{koe1998} observed a flattened H$\alpha$ core
and concluded that it is most likely due to the presence of a magnetic field
of $43\pm10$ kG. Assuming a dipole magnetic field, then the longitudinal
field can be at most 17 kG. We measured a longitudinal field of $3.4\pm5.0$ kG.
Within $2\sigma$ of the longitudinal field measurement the inclination has
to be greater $39^\circ$, however if we consider the uncertainty in the 
surface field measurement then we cannot constrain the inclination.

\paragraph{WD 2211-495:} This object was discussed in \citet{ven1999} who
placed an upper limit of 30 kG on the surface magnetic field from the narrow
H$\alpha$ core and assuming a simple dipole, the longitudinal field can be
at most 12 kG. We measured a longitudinal field of
$6.6\pm5.3$ kG and at $2\sigma$ we cannot place any tight constraints on the
inclination of the field.

\paragraph{WD 2359-434} An unusually narrow and flat H$\alpha$ core was 
reported by \citet{koe1998} and they speculated that a magnetic field could be 
the cause of this effect. A variable flattened core was also reported by
\citet{max1999}. A weak magnetic field was detected by \citet{azn2004}, who
measured a lower limit for the longitudinal field strength of 
$-4.504\pm0.958$ kG. Our measurement of  $3.4\pm4.4$ kG was not sensitive
enough to detect such a low magnetic field.
The trigonometric parallax from the Yale Parallax catalog implies a distance
of $7.8\pm0.4$ pc which is in agreement with the distance obtained from the
spectroscopic fit.

Table~\ref{tbl_survey1} also includes a number of stars that were not
observed using spectropolarimetry, but intensity spectra were obtained.
We placed an upper limit for the magnetic field of 1MG for these stars. We
assumed that the Zeeman splitting would be observed for white dwarfs with
surface fields higher than 1 MG.
Also we used these optical spectra to determine new effective temperatures
and surface gravities for these stars.

In addition, to the above discussed white dwarfs, there are a few objects
with peculiar properties that deserve discussion. 

\paragraph{WD 0141-675:} This is a known high proper-motion white dwarf, 
however few spectroscopic
observations of this star have been carried out. \citet{hol2002} listed this
white dwarf as local with a distance of 9.6 pc. Our spectroscopic studies
found this object to be a cool white dwarf ($T_{\rm eff} = 6460\pm160$ K)
with a distance of 9 pc.

\begin{figure}
\plotone{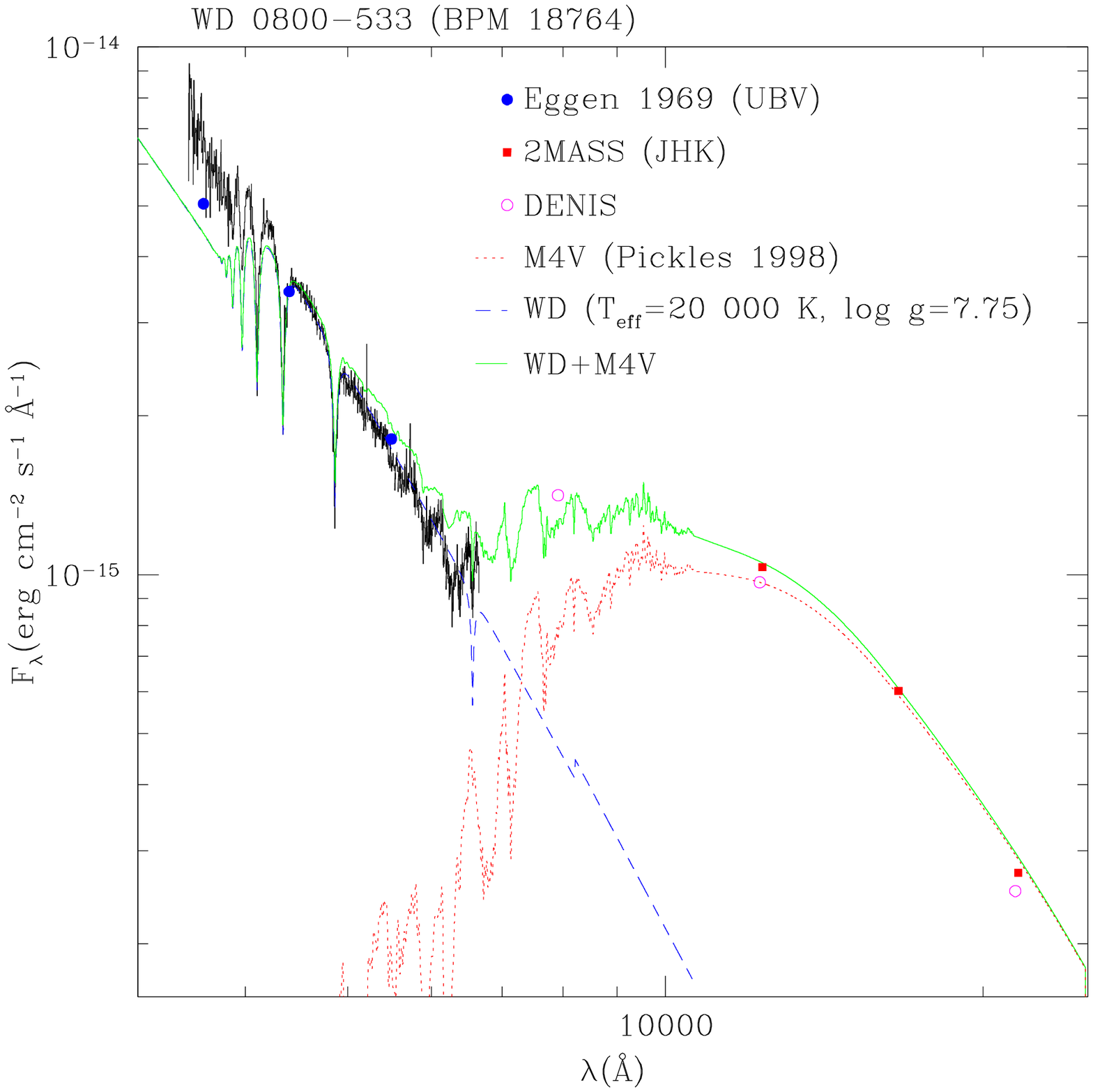}
\caption{Optical and infrared photometry compared to the observed spectrum of
WD 0800-533 and a white dwarf model spectrum ($T_{\rm eff} = 20000$ K and 
$\log{g} = 8.0$). Evidence for a cool companion is present with an infrared
excess.}
\label{fig_BPM18764}
\end{figure}

\paragraph{WD 0800-533:} This object was reported as a possible binary by
\citet{wic1977} who observed broad emission cores in H$\alpha$ and $H\beta$.
They also noted that it lies near the X-ray error box of 3U 0804-58. We
checked the ROSAT database for X-ray sources in the vicinity of this object,
however the closest object is about half a degree away, hence we conclude
WD 0800-533 is either not a strong X-ray source or its X-ray emission is 
variable.
Our spectra, which only covers the upper Balmer lines (H$\beta$-H8), 
did not show obvious signs of emission or a cool companion. However, we obtained
2MASS infrared (JHK) and DENIS (IJK)\footnote{Data available at http://cdsweb.u-strasbg.fr/CDS.html}
photometry and found that the white
dwarf has significant infrared excess which is possibly due to a cool companion.
We estimate the secondary to be a
M3-4V star by comparing a combined spectrum of a white dwarf and a M dwarf 
\citep{pic1998}, as shown in Figure~\ref{fig_BPM18764}. Further 
studies are required to determine its binary parameters.

\paragraph{WD 1223-659:} This is a local white dwarf ($d \sim 13$ pc) with
few spectroscopic observations. \citet{weg1973} reported weak Ca II lines, 
however our spectra only show weak hydrogen lines, in agreement with the
classification of \citet{wic1977}. Therefore, this object is a
cool DA white dwarf with $T_{\rm eff} = 7740\pm70$ K and 
$M=0.67\pm0.07\ M_\odot$.

\paragraph{WD 1236-495:} This is a well known massive ZZ Ceti star. We did not
detect the presence of a magnetic field down to about 10 kG. The Balmer lines
were fitted with model spectra to obtain an effective temperature of 
$11870\pm130$ K and a surface gravity of $\log{g} = 8.84\pm0.04$, and hence a 
mass of $1.11\pm0.02 M_\odot$.

\paragraph{WD 1628-873} has an effective temperature 
($T_{\rm eff} = 11160\pm190$ K) and a surface gravity ($\log{g} = 8.29\pm0.07$)
that place it near the instability strip. This star was checked for variability
by \citet{mcg1977} who found WD 1628-873 to be constant in luminosity, however
it is useful in helping define the instability strip.

\paragraph{WD 2159-754:} This is an ultra-massive white dwarf for which few
spectroscopic observations have been carried out. 
\citet{sch1981} determined an effective temperature of 
$T_{\rm eff} = 8700\pm400$ K and a surface gravity $\log{g}=8.1\pm0.2$ by
comparing the equivalent widths of H$\delta / \epsilon$ versus 
H$\beta / \gamma$. Our surface gravity of $\log{g}=8.95\pm0.09$ is 
significantly higher. 

\paragraph{WD 2336-079} was observed by \citet{kaw2004}. Their temperature and
surface gravity places it in the ZZ Ceti instability strip. \citet{gia2006} 
have observed this star and found it to be variable. We have re-analyzed 
this star using improved spectral models and found the effective temperature to 
be $T_{\rm eff} = 11010\pm210$ K and the surface gravity to be $\log{g} = 8.05$
which places this star on the red edge of the instability strip.

\paragraph{WD 2351-368:} This is a high proper-motion white dwarf, and the 
kinematics of this star make it a halo candidate
\citep{pau2003}. Halo white dwarfs are expected to be very cool and old, however
this star is quite hot ($T_{\rm eff} = 14540\pm320$ K) with an average mass
($M = 0.61\pm0.03 M_\odot$). \citet{pau2003} argue that this white dwarf 
evolved from a long-lived low mass star. This is one of the few stars for 
which we could not obtain a spectropolarimetric measurement, and therefore we 
can only conclude that if a magnetic field is present it must be less than 
$\sim 1$ MG.

\subsection{Ultramassive White Dwarfs}

As part of our general spectropolarimetric survey we have also observed 
EUV-selected ultra-massive ($\ge 1.1\ M_\odot$) white dwarfs, which are 
listed in Tables~\ref{tbl_bmeas} and \ref{tbl_survey1} and indicated with a 
tablenote. No magnetic fields were detected in these stars 
(except for the known magnetic white dwarf EUVE~J0823$-$25.4). In addition to 
the EUV-selected stars, two
ultramassive white dwarfs (WD 1236-495 and WD 2159-754) were observed as
part of the general survey and were found to be non-magnetic.

\begin{deluxetable}{lccc}
\tabletypesize{\scriptsize}
\tablecaption{Ultramassive White Dwarfs.\label{tbl_ultramass}}
\tablewidth{0pt}
\tablehead{
\colhead{Name} & \colhead{Mass} & \colhead{$B_p$} & \colhead{Reference} \\
\colhead{} & \colhead{($M_\odot$)} & \colhead{(MG)} & \colhead{} \\
}
\startdata
EUVE 0317-855 & $1.34 \pm 0.03$ & 450        & 2 \\
GD 50         & $1.27 \pm 0.01$ & $\la 0.12$ & 3,4 \\
EUVE 0653-564 & $1.16 \pm 0.06$ & $\la 0.27$ & 1 \\
EUVE 0823-254 & $1.27 \pm 0.06$ & 3.5        & 1,5 \\
EUVE 1024-303 & $1.21 \pm 0.05$ & $\la 0.12$ & 1 \\
LTT 4816      & $1.11 \pm 0.02$ & $\la 0.03$ & 1 \\
EUVE 1535-774 & $1.18 \pm 0.07$ & $\la 0.66$ & 1 \\
PG 1658+441   & $1.31 \pm 0.02$ & 3.5        & 6,5 \\
EUVE 1727-360 & $1.20 \pm 0.03$ & $\la 0.12$ & 1 \\
LTT 8816      & $1.17 \pm 0.03$ & $\la 0.03$ & 1 \\
\enddata
\tablerefs{(1) This work; (2) \citet{fer1997}; (3) \citet{ven1997};
(4) \citet{sch1995}; (5) \citet{fer1998}; 
(6) \citet{sch1992} }
\end{deluxetable}

Since the mass of magnetic white dwarfs is on average higher
($\sim 0.8\ M_\odot$) than the mass of non-magnetic white dwarfs
($\sim 0.6\ M_\odot$) we selected known ultra-massive white dwarfs for which
magnetic field measurements are available.
Table~\ref{tbl_ultramass} lists these
ultra-massive white dwarfs.
The presence of a magnetic field is not guaranteed in massive
white dwarfs but a higher incidence of magnetism is observed. For the white
dwarfs where a magnetic field has been detected, the previously published
values are given, otherwise an upper limit on the polar magnetic field from
this survey is provided. The upper limit is calculated from the measured 
$2\sigma$ error of the longitudinal component of the field and since these
upper limits are for reference only, we have assumed $57^\circ$ for the
inclination (i.e., the most probable angle:  $\int_0^{\pi/2} x \cos{x} dx$).

\begin{figure}
\plotone{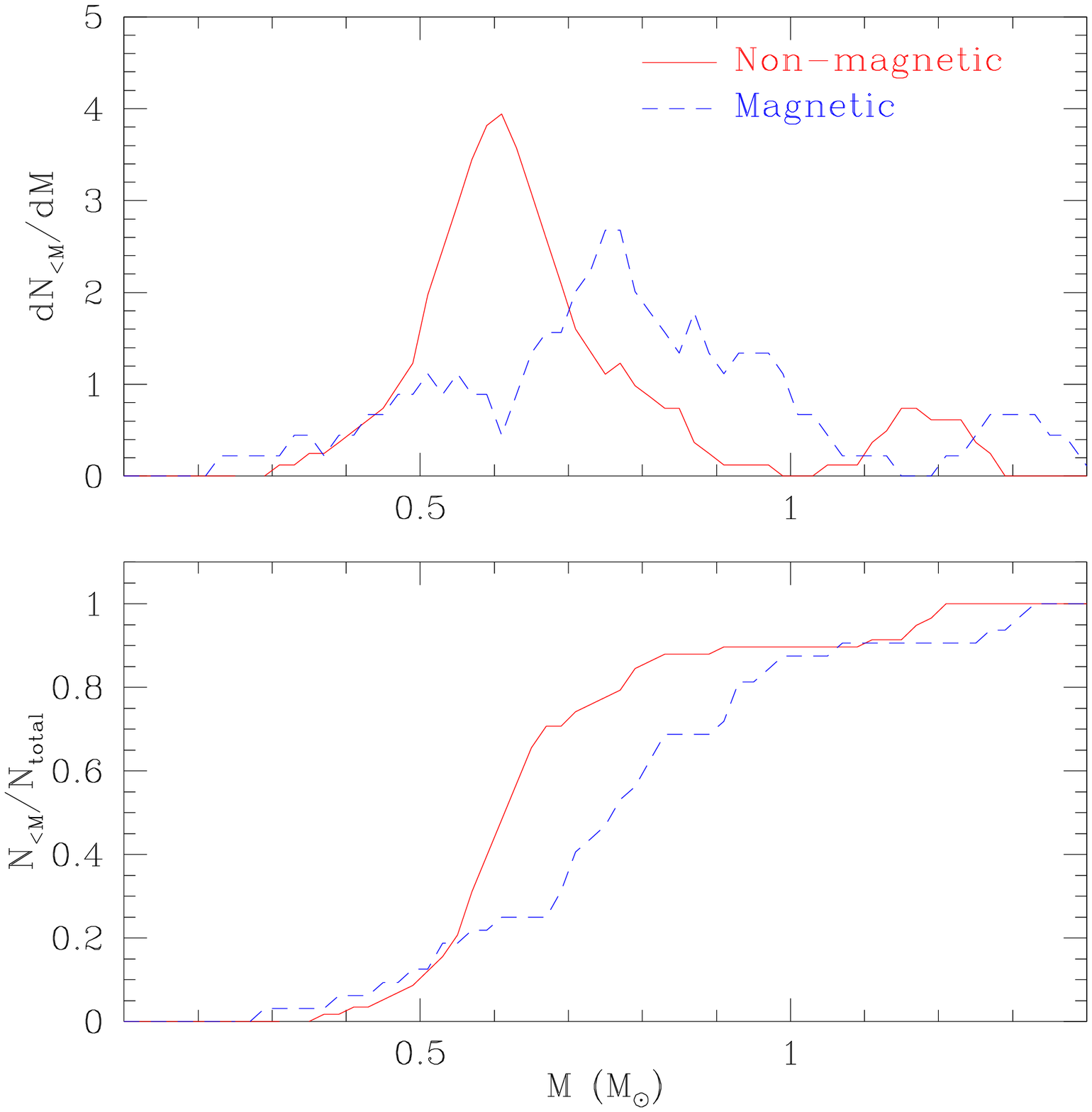}
\caption{Mass distribution of the non-magnetic (see text) and  
magnetic samples of white dwarfs.}
\label{fig_mass_dist}
\end{figure}

We have calculated the mass distribution of our sample of white dwarfs from
Table~\ref{tbl_survey1} (excluding the magnetic stars EUVE~0823$-$25.4 and 
WD~2359-434) 
and compared it to the mass distribution of magnetic white dwarfs, for which the
mass is known. These masses are given in Appendix A. The mass 
distributions and the cumulative distributions of the non-magnetic
and magnetic white dwarfs are shown in Figure~\ref{fig_mass_dist}. The
figure shows that magnetic white dwarfs do have higher masses than the 
non-magnetic white dwarfs. The mean mass of the magnetic white dwarf sample
is $\langle M \rangle = 0.78\ M_\odot$ with a dispersion of 
$\sigma = 0.24\ M_\odot$. The mode (i.e., the most probable mass) of the 
magnetic sample is $0.76\ M_\odot$. The mean mass of the non-magnetic 
sample is $\langle M \rangle = 0.68\ M_\odot$ with a dispersion of 
$\sigma = 0.20 M_\odot$. The mode of the non-magnetic sample is $0.57\ M_\odot$.
If we only consider the major peaks of the two distributions (i.e., excluding
the ultramassive white dwarfs), then the mean of the magnetic sample becomes
$\langle M \rangle = 0.73\ M_\odot$ with a dispersion of $\sigma = 0.19 M_\odot$
and the mean of the non-magnetic sample becomes 
$\langle M \rangle = 0.62\ M_\odot$ with a dispersion of 
$\sigma = 0.11 M_\odot$. \citet{wic2005} suggest that the mass distribution
is naturally biased toward a higher mass since they assumed high-field magnetic 
white dwarfs evolve from more massive stars. And the low-field white dwarfs are
assumed to evolve from low-mass main-sequence stars which will produce
a mass distribution which is similar to that of non-magnetic white dwarfs. 

\subsubsection{EUVE J0823-25.4}

\begin{figure}
\plotone{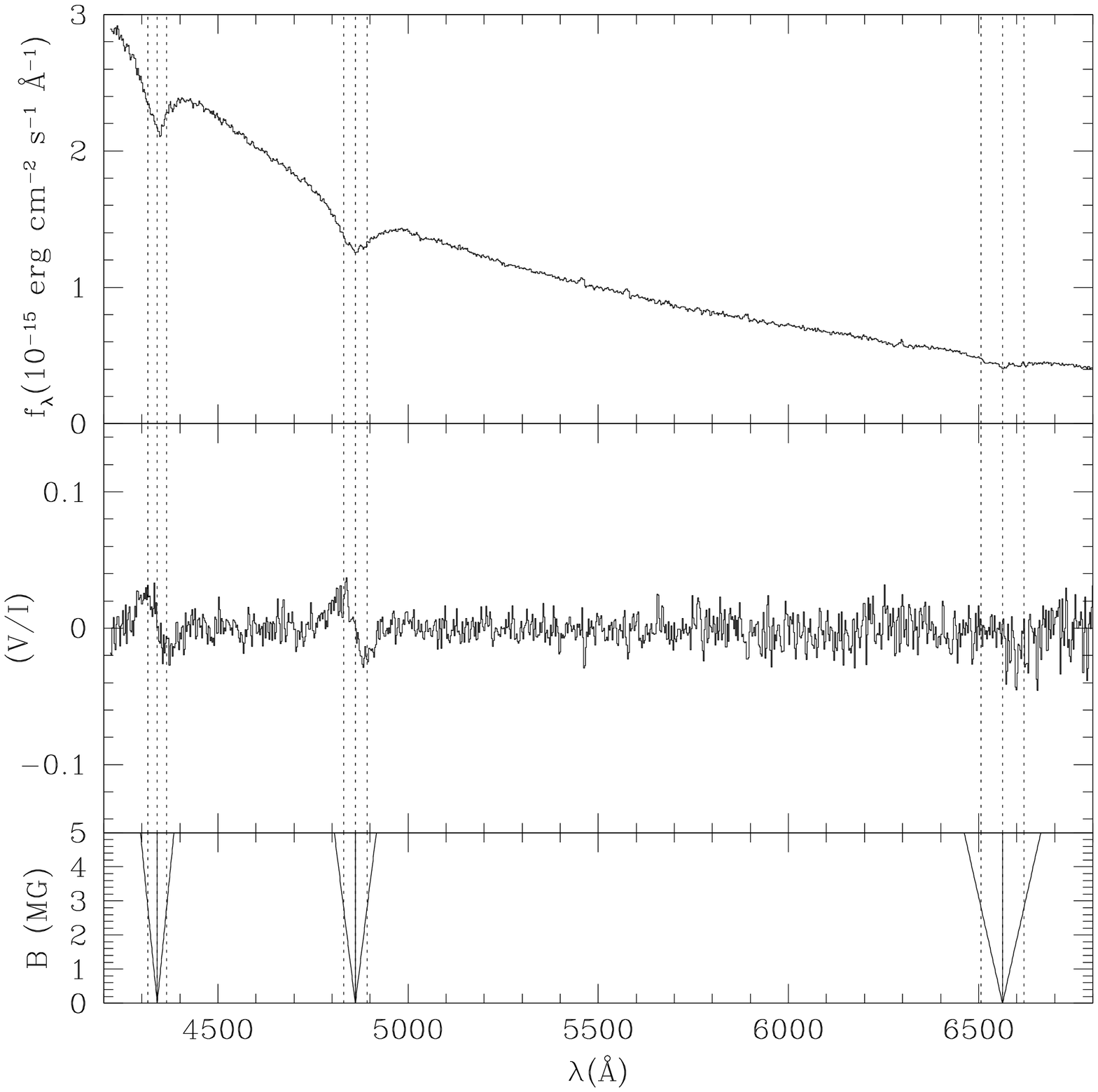}
\caption{The flux (top) and circular polarization (bottom) spectra of the
ultra-massive magnetic white dwarf EUVE~J0823$-$25.4.}
\label{fig_EUV0823}
\end{figure}

We also observed the massive white dwarf EUVE J0823$-$25.4, shown in
Figure~\ref{fig_EUV0823}, which is a known magnetic white dwarf.
\citet{fer1998} measured a dipole field of 3.5 MG inclined at a viewing angle
of $60^\circ$ using the Zeeman split Balmer line profiles. We
measured a longitudinal field of $\sim\ -600$ kG. Using the
relationship $B_p = B_\ell/(0.4 \cos i$), and B$_p = 3.5$ MG, the
inclination would be $\sim 65^\circ$, which is in agreement with
\citet{fer1998}.

The measurement of the
magnetic field at H$\alpha$ resulted in  a lower value of $B_\ell \sim 63$ kG
compared to $\sim 600$ kG for H$\beta$ and H$\gamma$. The H$\alpha$ line
profile is dominated by the quadratic Zeeman effect, and the linear Zeeman
approximation assumed by the measurement technique is no longer valid and was
therefore excluded in calculating the longitudinal field.

\subsection{Binary Stars}

Table~\ref{tbl_survey_bin} shows the close binaries containing a white dwarf
that were observed as part of the survey, giving their apparent magnitude,
spectral type, effective temperature and surface gravity.

The spectropolarimetry of BPM 6502 (WD 1042$-$690) suggests the presence of a
weak magnetic
field. However, because the system is a close binary the Balmer line profiles
are shifted between different waveplate exposures.
When the final spectra, which have opposite polarization, are subtracted from
one another, then the shifted Balmer line profiles can cause the same effect
as the shifted $\sigma$ components of the Zeeman effect. Therefore the
longitudinal field measurements of close binary systems need to be viewed with
caution. \citet{azn2004} also observed this system and did not detect the
presence of a magnetic field.
A similar effect appears to occur for EUVE J0720$-$317.
For the close binary LTT 1951, only H$\beta$ and H$\gamma$ were used in the
measurement of magnetic field strength. The cool companion
dominates the spectrum in the red and therefore H$\alpha$ could not be used
in the measurement of the magnetic field of the white dwarf. The measurements
we obtained for  LTT 1951 indicate that the white dwarf does not possess a
strong magnetic field.

Magnetic white dwarfs in binary systems have only been observed in
cataclysmic variables or in double degenerate systems such as EUVE~0317$-$855
\citep{fer1997} and EUVE~1439$+$750 \citep{ven1999b}. The distribution of
magnetic field strengths of magnetic white dwarfs in cataclysmic variables
appears to be similar to the distribution of $B$ of isolated magnetic white
dwarfs \citep{wic2000}. However, there appears to be a paucity of high-field
strengths in white dwarfs in cataclysmic variables, which may be a selection
effect \citep{wic2000}.
Many magnetic cataclysmic variables are known but no post-common
envelope binaries are known to contain a magnetic white dwarf, with the 
possible exception of SDSS~J121209.31+013627.7 \citep{sch2005} which was found
to be a magnetic white dwarf (B$_p = 13$ MG) with a probable brown dwarf 
companion with an orbital period of $\sim 90$ minutes. 
\citet{lie1995} suggested that the reason for the non-detection of magnetic
fields in post-common envelope binaries may be due to selection effects,
for example the contamination by the secondary spectral features may
hide the features that would identify the white dwarf as magnetic.
Note that our spectropolarimetric survey of white dwarfs in close binaries
with red dwarfs should have easily allowed detection of a field typical in 
magnetic accretors.

\subsection{Non-DA stars}

\begin{figure}
\plotone{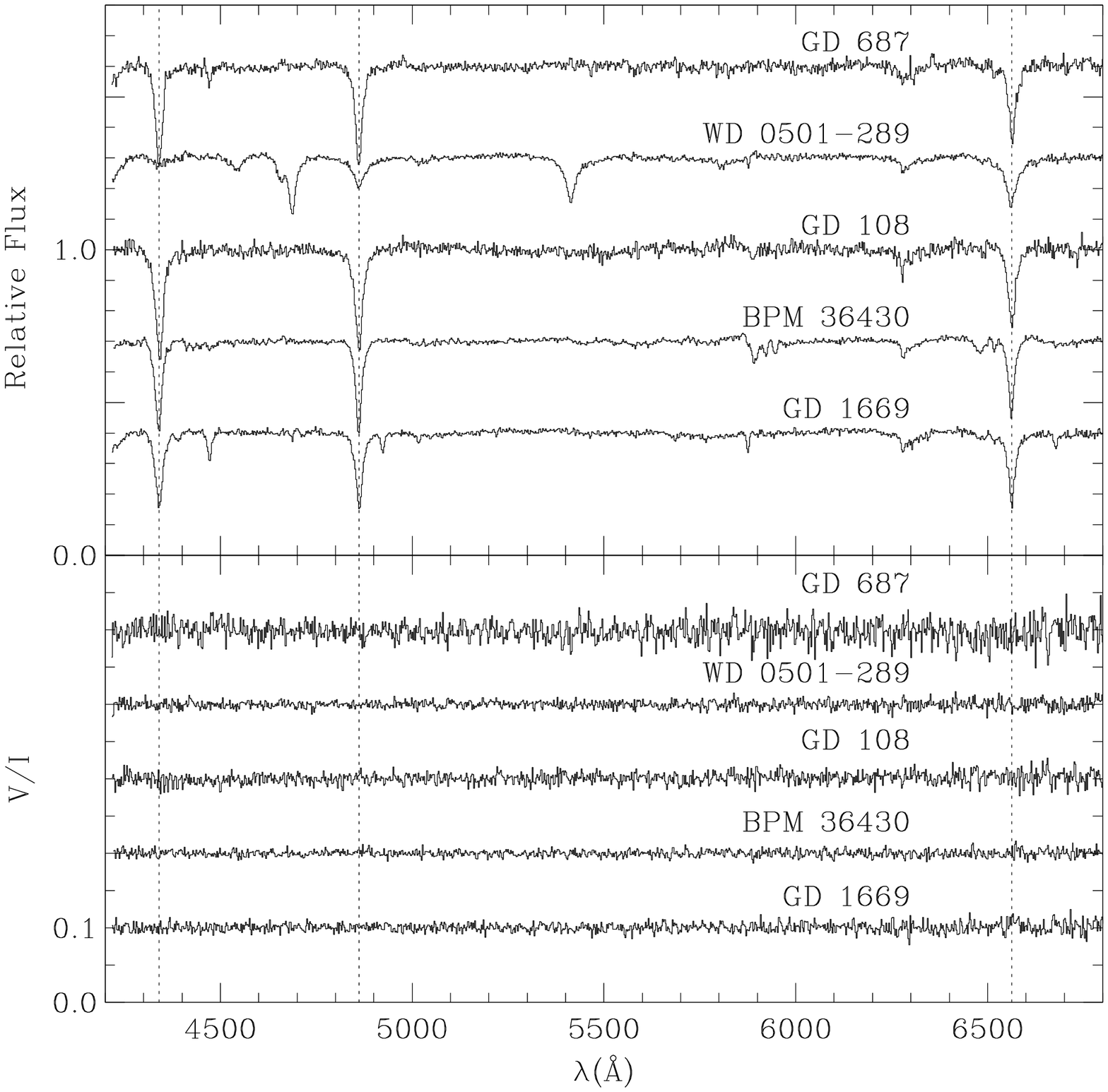}
\caption{Flux and circular polarization spectra of the four subdwarfs and 
one DO white dwarf, which are given in Table~\ref{tbl_survey_oth}.)}
\label{fig_specpol3}
\end{figure}

As part of our survey we also observed stars that are non-DA, four of these
are sdB stars and one is a DO white dwarf. 
Table~\ref{tbl_survey_oth} lists these stars and Figure~\ref{fig_specpol3} shows
their flux and polarization spectra. We did not detect the presence of a 
magnetic field in any of these stars.

Two of these stars were misclassified as DA white dwarfs, however the spectra
of BPM 36430 and GD 1669 \citep[see also][]{bee1992,lis2005} show them to be
sdB stars. GD 1669 (WD 2329$-$291) is believed to be magnetic. \citet{koe1998} 
observed a broadened H$\alpha$ core in GD 1669 and concluded that it is likely 
due to presence of a magnetic field of $31\pm10$ kG. Therefore an upper limit 
on the longitudinal field would be 12.4 kG, and our measurements 
($0.2\pm5.8$ kG and $7.1\pm6.7$) clearly lie within this limit. Within $2\sigma$
of the measurements we cannot place any tight constraints on the inclination
if a magnetic field of 31 kG is present.
Until recently no magnetic fields have been
known to exist in sdB stars. \citet{oto2005} have searched for magnetic fields
in six sdB/O stars, which resulted in one clear detection of magnetism in
HD 76431 and marginal detections in the remaining 5 stars. The 
broadened Balmer core reported by \citet{koe1998} may indicate a weak field or 
it may also suggest that this star is a possible binary (sdB + WD/sdB).
More observations of GD 1669 are necessary to determine its status.

For three of the subdwarf B stars effective temperatures and surface gravities
have been published, however for BPM 36430, no temperature and surface 
gravities were found in the available literature. We calculated a grid of LTE
line-blanketed spectra for temperatures ranging from 16000 K to 40000 K (in 
steps of 4000 K), surface gravities between 4.5 to 7.0 (in steps of 0.25 dex) 
and He-abundances of $\log{(N_{He}/N_{H})}$ = -4 to 0 (in steps of 0.5). 
We determined the effective temperature, surface gravity and He-abundance for
BPM 36430 by fitting the Balmer lines (excluding H$\alpha$) and He lines with
synthetic spectra. We found $T_{\rm eff} = 30080\pm660$ K,
$\log{g} = 5.15\pm0.16$ and $\log{(N_{He}/N_{H})} \la -3$. The spectrum of 
BPM~36430 also exhibits NaI and CaII lines.
Note that \citet{nap1997}
found that for sdBs, LTE models begin to deviate from NLTE models above
30000 K, therefore the temperature and gravity of BPM 36430 are probably only 
slightly lower than our LTE determination.

The magnetic field for the DO white dwarf MCT 0501$-$2858 was measured using 
4 helium lines, HeII 4686, HeII 4859, HeII 5412 and HeII 6560.
We measured a longitudinal magnetic field of $-5.6\pm10.4$ kG indicating that
the white dwarf does not have a strong magnetic field. 

\section{Local population of magnetic white dwarfs}

\citet{hol2002} lists 46 white dwarfs that reside within 13 pc of the Sun,
however three of these were found to be F-type stars by \citet{kaw2004} and
need to be removed from the list. 
Therefore there are 43 known white dwarfs
within 13 pc of the Sun of which 9 are magnetic resulting in an incidence
of $21\pm8\%$. Similarly, \citet{hol2002} lists 109 white dwarfs residing
within 20 pc of the Sun, apart from the three stars already mentioned, the
white dwarf WD 1717-345 should be excluded from the list since its distance
places it at $\sim 150$ pc. We can add 9 additional stars to the list, eight
from \citet{kaw2006} and the newly discovered cool white dwarf PM~J13420-3415
from \citet{lep2005}. Also \citet{zuc2003} found that WD0532+414 and WD0322-019
are double degenerate binaries, therefore we have a total of 116 white dwarfs
within 20 pc of the Sun\footnote{The double degenerate classification of these
objects means that they are most likely further away than 20 pc, however since 
the nature of the secondary stars are unknown we cannot be sure of their total
luminosity and will use them in our calculations.}.
Of these 15 are classified as magnetic resulting in an incidence
of $13\pm4\%$. The properties of all the known magnetic white dwarfs to date 
are given in Appendix A.

\begin{figure}
\plottwo{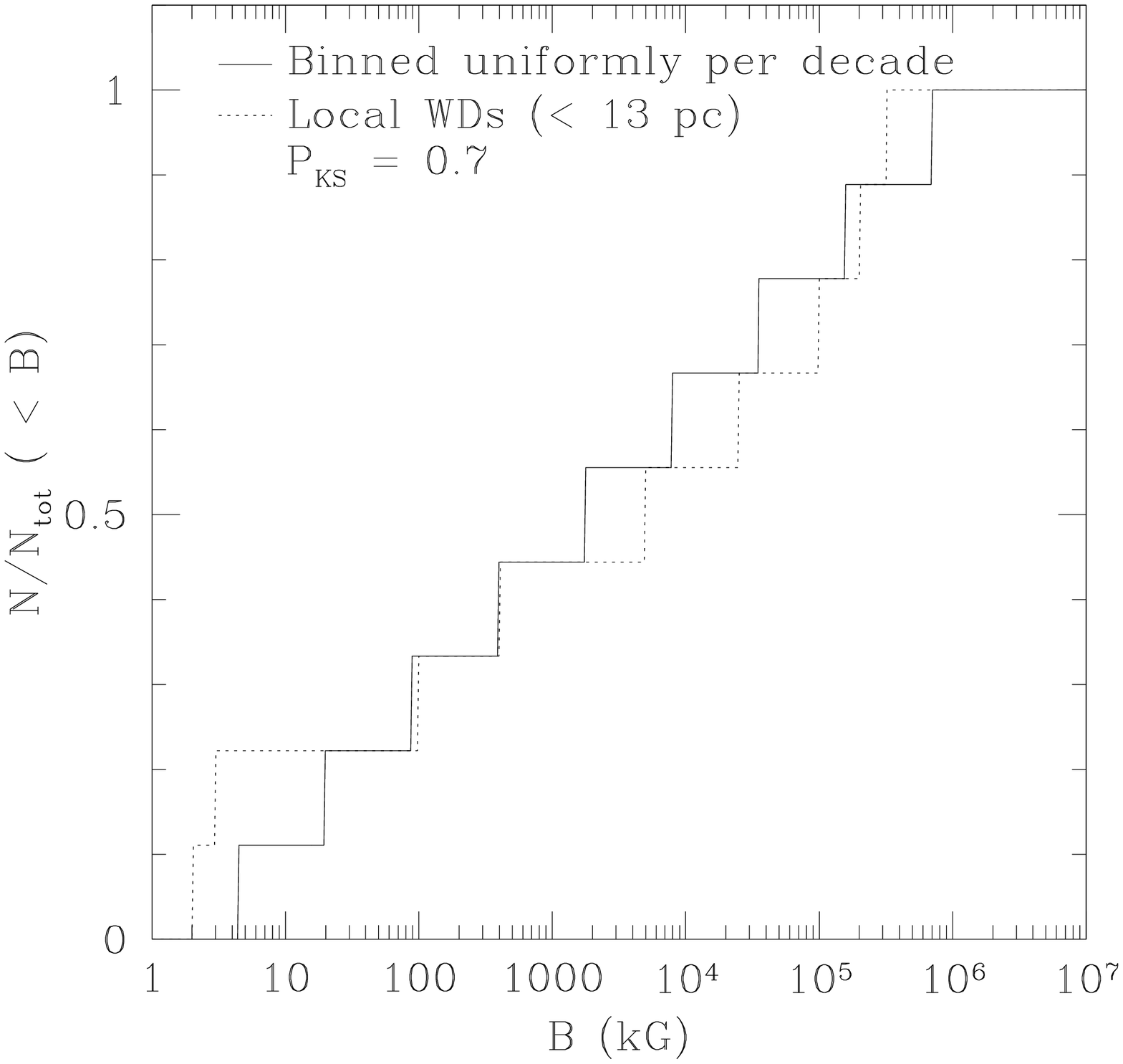}{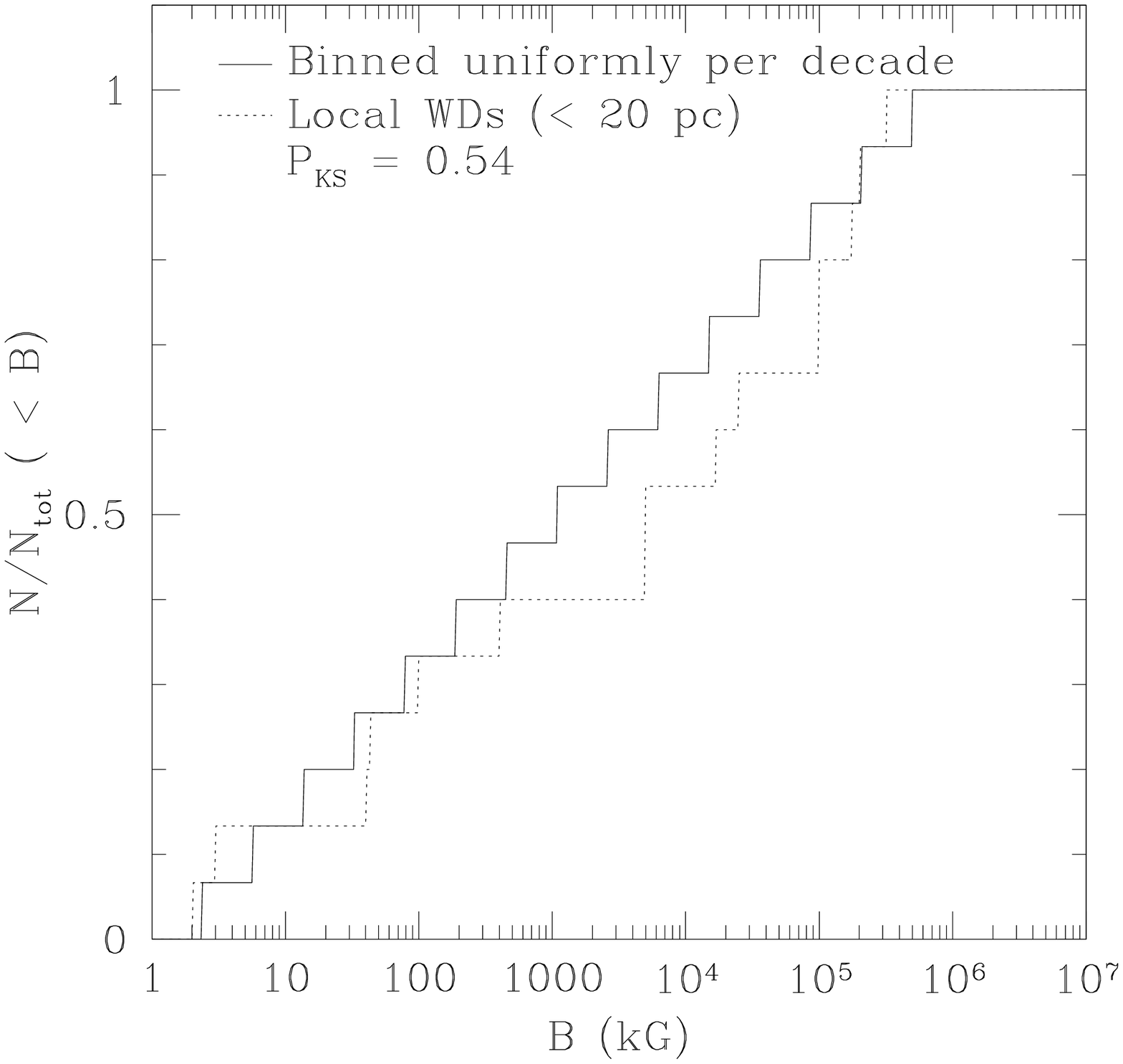}
\caption{Cumulative distributions of magnetic white dwarfs which are with
13 pc ({\it left}) and 20 pc ({\it right}) of the Sun, compared to a
distribution with a constant number per decade interval.}
\label{fig_local}
\end{figure}

Figure~\ref{fig_local} shows the cumulative distribution
of magnetic field strengths of white dwarfs found within 13 and 20 pc. A 
Kolmogorov-Smirnov (KS) test for the white dwarfs within 13 pc shows that the
incidence of magnetic white dwarfs appear constant for each decade
interval with a probability of 0.7. For the white dwarfs within 20 pc,
this probability is reduced to 0.53. The discovery of white dwarfs with
kG fields by \citet{azn2004} could imply that there is a significant population
of white dwarfs with weak magnetic fields.

We have searched the literature for all 116 white dwarfs within 20 pc to find 
out to what level of magnetic field strength sensitivity each white dwarf
has been observed. We have distributed the white dwarfs into four bins which
are based on the sensitivity achieved during the observations.
Table~\ref{tbl_local} provides the number of magnetic white dwarfs known in 
each bin as well as the number of white dwarfs that have been checked for 
magnetic fields with the corresponding strength for that bin. 

\begin{deluxetable*}{lccccc}
\tabletypesize{\scriptsize}
\tablecaption{Statistics of magnetism in local white dwarfs.\label{tbl_local}}
\tablewidth{0pt}
\tablehead{
\colhead{$B_p$} & \colhead{No. of Magnetic WDs} & \colhead{No. of non-magnetic WDs} & \colhead{Fraction} \\
}
\startdata
$\ge$ 10 MG    & 8 & 14 & 8/107 \\
1MG - 10 MG  & 1 & 31 & 1/85 \\
100kG - 1 MG & 2 & 18 & 2/53 \\
$<$ 100 kG   & 4 & 29 & 4/33 \\
\enddata
\end{deluxetable*}

Based on these results, we have calculated probabilities in finding more 
magnetic white dwarfs within 20 pc. In our calculations we have assumed that 
the $B_p$ value of a given bin is a lower limit 
to which the white dwarfs have been checked, i.e., if a white dwarf has been 
checked for magnetism down to 1 MG it has also been checked for fields larger
than the next bin at 10 MG. We have calculated the probability of finding more 
magnetic white dwarfs within a given sensitivity bin assuming a binomial 
probability distribution, i.e., 
\begin{equation}
P(X=x)={n\choose x} p^x (1-p)^{n-x}
\end{equation}
We calculate the probability ($P$) of finding $x$ number of magnetic white 
dwarfs in the total number of white dwarfs $n$ that have not been checked
for magnetism at the sensitivity level of the given bin. The value of $p$ is 
determined from the fraction of magnetic white dwarfs in a given bin out of 
the total number of white dwarfs that have been checked for magnetism in that
bin.

In the first bin, 107 out 116 white dwarfs
have been checked for magnetic fields greater than or equal to 10 MG. The 
remaining 9 objects are mostly DC white dwarfs, with one DQ (WD 1043-188). 
WD~1132-325 (LHS 309) may also be a DQ white dwarf rather than a DC 
\citep{hen2002}. There are 8 white dwarfs with measured $B_p \ge 10$ MG.
Continuum polarization should be detectable in all white dwarfs
with $B_p \ge 10$ MG. For white dwarfs which display H or
He lines, then these lines will be significantly displaced by magnetic fields
stronger than 10 MG. Therefore based on the fraction of white dwarfs with 
$B_p \ge 10$ MG ($8/107 = 0.075$), there is a 50\% probability of finding at 
least one white dwarf with $B_p \ge 10$ MG, in the 9 stars.

In the next bin, we have only one magnetic white dwarf (WD 0548-001)
and 31 stars that have been checked for magnetic fields between 1 MG
and 10 MG. In this bin, white dwarfs showing either H or He lines will 
display Zeeman splitting and therefore good quality spectroscopy will be 
sufficient to identify white dwarfs with 1 MG $\le B_p < 10$ MG.
And we have 23 stars that are not classified as magnetic and have
not been checked for magnetism between 1 MG and 10 MG. Therefore, the 
fraction of white dwarfs with 1 MG $\le B_p <$ 10 MG is 1/85. Assuming that this
fraction represents the probability of finding a magnetic white dwarf with
a field of 1 MG $\le B_p <$ 10 MG, then the probability that at least one of 
the 23 white dwarfs (not checked for fields down to 1 MG level) is 0.24. 
Therefore, it is unlikely that more white dwarfs will be found with 
1 MG $\le B_p <$ 10 MG.

In the 100 kG $\le B_p <$ 1 MG bin, again we have two magnetic white dwarfs
(WD~0009+501 and WD~0728-642) and 53 stars that have been checked for magnetic 
fields in this bin. There are 54 stars that are not classified magnetic and 
have not been checked for magnetism between 100 kG and 1 MG. The fraction of 
white dwarfs with 100 kG $\le B_p <$ 1 MG is 2/53 and the probability of 
finding at least one magnetic white dwarf with a field between 100 kG and 1 MG 
is 0.87 and is therefore a very likely eventuality.

In the final bin ($B_p \le$ 100 kG), we have 4 magnetic white dwarfs 
(WD~0413-077, WD~1953-011, WD~2105-820 and WD~2359-434) out of 33 stars that 
have been checked for magnetism at this level. There we can assume that 12\% 
of white dwarfs have magnetic fields below 100 kG. A lower limit to this bin 
is probably 10 kG, based on the sensitivity of the surveys taken to measure the 
magnetic field
strengths in the 33 stars. There are 72 stars that are not classified as
magnetic and that have not been checked for magnetic fields below 100 kG.
Note that some of these stars are likely to be in the SPY sample of white 
dwarfs \citep{nap2001} where magnetic fields less than 100kG would be 
detectable in the core of the Balmer lines, however since many of the spectra 
are not yet published we cannot include them in the analysis. 
The probability that at least one of these stars has a magnetic field less
than 100 kG is almost 1. If we calculate the probability of finding a 
particular number of magnetic white dwarfs in this sample, then the
probability peaks at finding 8 white dwarfs $B_p \le 100$ kG among the 72 
(P = 0.14). Therefore it is very likely that there are many white dwarfs within 
20 pc that have magnetic fields less than 100 kG and the flat distribution
of field per decade is probably fortuitous at these low fields.

\begin{figure}
\plotone{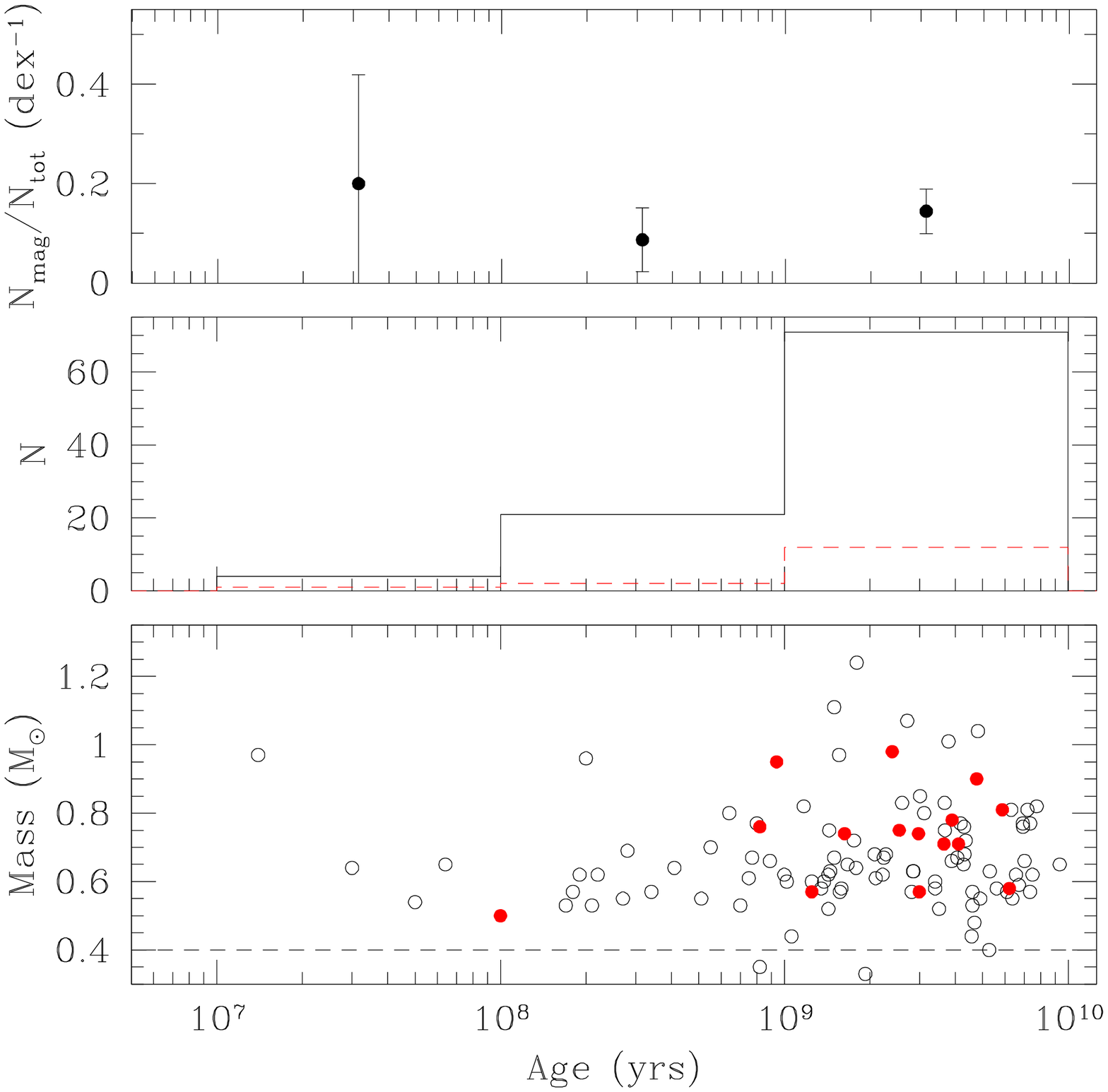}
\caption{Masses and ages for magnetic and non-magnetic white dwarfs within 20pc 
of the Sun, which were interpolated using relations described in \S 3.}
\label{fig_age_mass}
\end{figure}

For all the white dwarfs within 20 pc we have obtained the effective 
temperature, mass and cooling age from the literature. Those for which we could
not find a calculated cooling age, we calculated the cooling age using the
evolutionary models of \citet{ben1999}. For a few white dwarfs we assumed
a surface gravity of $\log{g} = 8.0$ and hence a mass of $0.57 M_{\odot}$.
A plot of the mass versus the age for these white dwarfs is shown in
Figure~\ref{fig_age_mass}. The filled circles are the known magnetic white 
dwarfs within 20 pc and the open circles are the non-magnetic white dwarfs
within 20 pc. The plot includes the suspected magnetic white dwarf WD 2105-820
\citep{koe1998}.
Figure~\ref{fig_age_mass} also shows the distribution of the cooling age
binned into decades, which shows that the local population is relatively old
with most white dwarfs having a cooling age greater than $10^9$ years.
We also calculated the fraction of magnetic white dwarfs for each decade 
interval of age, which is shown in the top panel of Figure~\ref{fig_age_mass}.
From this plot we can see that there does not appear to be a higher incidence
of magnetism among older white dwarfs within the uncertainties of the fractions
in each bin. This is in contradiction to other similar studies, such as
\citet{lie2003} and \citet{val1999}, which found
that the incidence of magnetism is higher in older white dwarfs. In terms of 
absolute fractions, this appears to be true, but due to the large uncertainty 
in hot white dwarfs, it is not conclusive, and magnetic incidence may be
constant as a function of temperature.

\section{Summary}

We have conducted a spectropolarimetric survey of southern white dwarfs, which
resulted in no new detections of magnetic fields with the possible exception of
WD~0310-688 for which we measured a longitudinal field of $-6.1\pm2.2$ kG. 
We have also observed the known magnetic white dwarf EUVE~J0823-25.4 and 
measured a longitudinal field of $-600$ kG.
However, this survey has
helped constrain the incidence of magnetism in the Solar neighborhood. We
reviewed the list of known white dwarfs in the Solar neighborhood and found
that $21\pm8 \%$ are magnetic. We also investigated the probability of finding 
more magnetic white dwarfs within the Solar neighborhood based on the current
magnetic white dwarf incidence. We found
that a significant number of magnetic white dwarfs with fields $B < 100$ kG,
remain to be detected within the Solar neighborhood.

\begin{acknowledgements}
We thank Mount Stromlo Observatory for the generous allocation of time on the
74 inch telescope. We also wish to thank K. Vanlandingham, for assistance in
compiling the list of magnetic white dwarfs contained in the appendix.
A. Kawka acknowledges support from the Murdoch Student Fellowship and 
the Grant Agency of the Czech Republic (GA \v{C}R) 205/05/P186. S. Vennes 
acknowledges support from the College of
Science, Florida Institute of Technology. This research was supported by
Australian National University research grant. Studies of magnetic stars and
stellar systems at Steward Observatory is supported by the NSF through grant
AST 97-30792.
This publication makes use of data products from the Two Micron All Sky Survey, 
which is a joint project of the University of Massachusetts and the Infrared 
Processing and Analysis Center/California Institute of Technology, funded by 
the National Aeronautics and Space Administration and the National Science 
Foundation.

\end{acknowledgements}

\appendix
\section{Known Magnetic White Dwarfs\label{appendixA}}

Table~\ref{tbl_knownMWD} lists all the known magnetic white dwarfs as of 
June 2006. The table gives the WD number, an alternate name, the surface 
composition of the white dwarf (i.e., whether it is H- or He-rich), the polar 
magnetic field strength,
the effective temperature, the mass of the white dwarf if known, the rotational
period if known and references. For a number of white dwarfs, the properties
such as the effective temperature and magnetic field strength were determined
assuming a surface gravity of $\log{g} = 8.0$. For these white dwarfs this is
explicitly labeled in the Mass column instead of providing a mass 
determination.

In Table~\ref{table_non_mag_wd} we list the white dwarfs which have once been 
classified as magnetic due to their
peculiar spectra but which have since been shown to be non-magnetic. The
references that show the white dwarf not to be magnetic are provided. Note that
the stars are considered non-magnetic at the level advertised in the
literature and low-magnetic field in order of kG may still be present in these 
stars.

\clearpage

\LongTables
\begin{landscape}
\begin{deluxetable}{llrrrrrl}
\tabletypesize{\scriptsize}
\tablecaption{Known Magnetic White Dwarfs.\label{tbl_knownMWD}}
\tablewidth{0pt} 
\tablehead{
\colhead{WD} & \colhead{Other Names} & \colhead{Comp.} & \colhead{$B_p$} & \colhead{$T_{\rm eff}$} & \colhead{M}  & \colhead{$P_{rot}$} & \colhead{References} \\
\colhead{}   & \colhead{}            & \colhead{}      & \colhead{(MG)}  & \colhead{(K)}           & \colhead{($M_\odot$)} & \colhead{} & \colhead{} \\
}
\startdata
0003$-$103 & SDSS J000555.91$-$100213.4\tablenotemark{a} & He/C & ?    & 29000          & \nodata         & \nodata      & 1,2  \\
0009$+$501 & LHS 1038                   & H        & $\la 0.2$         & $6540\pm150$   & $0.74\pm0.04$   & $2-20$ hr    & 3,4 \\
0011$-$134 & LHS 1044                   & H        & $16.7\pm0.6$      & $6010\pm120$   & $0.71\pm0.07$   & \nodata      & 5,4 \\
0015$+$004 & SDSS J001742.44$+$004137.4 & He       & 8.3               & 15000          & \nodata         & \nodata      & 1  \\
0018$+$147 & SDSS J002129.00$+$150223.7 & H        & 550               & 7000           & \nodata         & \nodata      & 6  \\
0040$+$000 & SDSS J004248.19$+$001955.3\tablenotemark{b} & H & 14      & 11000          & \nodata         & \nodata      & 1  \\
0041$-$102 & Feige 7                    & H/He     & 35                & 20000          & $\log{g}=(8.0)$ & 131.606 min  & 7,8 \\     
0140$+$130 & SDSS J014245.37$+$131546.4 & He       & 4                 & 15000          & \nodata         & \nodata      & 1  \\
0155$+$003 & SDSS J015748.15$+$003315.1 & He (DZ)  & 3.7               & $\sim 6000$    & \nodata         & \nodata      & 1  \\
0159$-$032 & MWD 0159$-$032             & H        & 6                 & 26000          & log g = (8.0)   & \nodata      & 9 \\
0208$+$002 & SDSS J021116.34$+$003128.5 & H        & 490               & 9000           & \nodata         & \nodata      & 1  \\
0209$+$210 & SDSS J021148.22$+$211548.2 & H        & 210               & 12000          & \nodata         & \nodata      & 6  \\
0233$-$083 & SDSS J023609.40$-$080823.9 & H(DQA)   & 5                 & 10000          & \nodata         & \nodata      & 6  \\
0236$-$269 & HE 0236$-$2656\tablenotemark{c} & He  & ?                 & $6000-7000$    & \nodata         & \nodata      & 10 \\
0253$+$508 & KPD 0253$+$5052            & H        & $13-14$           & 15000          & $\log{g}=(8.0)$ & \nodata      & 11,12 \\
0257$+$080 & LHS 5064                   & H        & $\sim 0.1$        & $6680\pm150$   & $0.57\pm0.09$   & \nodata      & 4 \\
0301$-$006 & SDSS J030407.40$-$002541.7 & H        & 10.8              & 15000          & \nodata         & \nodata      & 13 \\
0307$-$428 & MWD 0307$-$428             & H        & 10                & 25000          & $\log{g}=(8.0)$ & \nodata      & 9 \\
0325$-$857 & EUVE~J0317$-$855           & H        & $185-450$         & 33000          & 1.35            & 725 s        & 14 \\
0329$+$005 & KUV 03292$+$0035           & H        & 12.1              & 26500          & \nodata         & \nodata      & 13 \\
0330$-$000 & HE 0330$-$0002\tablenotemark{b} & He  & ?                 & $6000-7000$    & \nodata         & \nodata      & 10 \\
0340$-$068 & SDSS J034308.18$-$064127.3 & H        & 45                & 13000          & \nodata         & \nodata      & 1 \\
0342$+$004 & SDSS J034511.11$+$003444.3 & H        &  1.5              & 8000           & \nodata         & \nodata      & 13 \\
0413$-$077 & 40 Eri B                   & H        & $0.0023\pm0.0007$ & $16490\pm84$   &$0.497\pm0.005$  & \nodata      & 15,16 \\
0446$-$789 & BPM 3523                   & H        & 0.00428           & $23450\pm20$   & $0.49\pm0.01$   & \nodata      & 17  \\
0503$-$174 & LHS 1734                   & H        & $7.3\pm0.2$       & $5300\pm120$   & $0.37\pm0.07$   & \nodata      & 5,4 \\
0548$-$001 & G 99$-$37                  & C$_2$/CH & $\sim 10$         & $6070\pm100$   & $0.69\pm0.02$   & 4.117 hr     & 18,19,20 \\
0553$+$053 & G 99$-$47                  & H        & $20\pm3$          & $5790\pm110$   & $0.71\pm0.03$   & 0.97 hr      & 21,4,20 \\
0616$-$649 & EUVE~J0616$-$649           & H        & 14.8              & 50000          & $\log{g}=(8.0)$ & \nodata      & 22 \\
0637$+$477 & GD 77                      & H        & $1.2\pm0.2$       & $14870\pm120$  & 0.69            & \nodata      & 23,24 \\
0728$+$642 & G 234$-$4                  & H        & $0.0396\pm0.0116$\tablenotemark{d} & $4500\pm500$ & \nodata & \nodata & 25 \\
0745$+$304 & SDSS J074850.48$+$301944.8 & H        & 10                & 22000          & \nodata         & \nodata      & 6  \\
0755$+$358 & SDSS J075819.57$+$354443.7 & H        & 27                & 22000          & \nodata         & \nodata      & 1 \\
0756$+$437 & G 111$-$49                 & H        & 220               & $8500\pm500$   & \nodata         & \nodata      & 26,25 \\
0801$+$186 & SDSS J080440.35$+$182731.0 & H        & 49                & 11000          & \nodata         & \nodata      & 6  \\
0802$+$220 & SDSS J080502.29$+$215320.5 & H        & 5                 & 28000          & \nodata         & \nodata      & 6  \\
0804$+$397 & SDSS J080743.33$+$393829.2 & H        & 49                & 13000          & \nodata         & \nodata      & 1 \\
0806$+$376 & SDSS J080938.10$+$373053.8 & H        & 40                & 14000          & \nodata         & \nodata      & 6  \\
0814$+$043 & SDSS J081648.71$+$041223.5 & H        & 10:               & 11500          & \nodata         & \nodata      & 6  \\
0816$+$376 & GD 90                      & H        & 9                 & 14000          & $\log{g}=(8.0)$ & \nodata      & 27,25 \\
0821$-$252 & EUVE~J0823$-$254           & H        & $2.8-3.5$         & $43200\pm1000$ & $1.20\pm0.04$   & \nodata      & 28 \\
0825$+$297 & SDSS J082835.82$+$293448.7 & H        & 30                & 19500          & \nodata         & \nodata      & 6  \\
0837$+$199 & EG 61                      & H        & $\sim 3$          & $17100\pm350$  &$0.817\pm0.032$  & \nodata      & 29 \\
0837$+$273 & SDSS J084008.50$+$271242.7 & H        & 10                & 12250          & \nodata         & \nodata      & 6  \\
0839$+$026 & SDSS J084155.74$+$022350.6 & H        & 6                 & 7000           & \nodata         & \nodata      & 1 \\
0843$+$488 & SDSS J084716.21$+$484220.4\tablenotemark{b} & H & $\sim 3$ & 19000         & \nodata         & \nodata      & 1 \\
0853$+$163 & LB 8915                    & H/He     & $0.75-1.0$        & 24000          & $\log{g}=(8.0)$ & \nodata      & 30 \\
0855$+$416 & SDSS J085830.85$+$412635.1 & H        & 2                 & 7000           & \nodata         & \nodata      & 1  \\
0903$+$083 & SDSS J090632.66$+$080716.0 & H        & 10                & 17000          & \nodata         & \nodata      & 6  \\
0904$+$358 & SDSS J090746.84$+$353821.5 & H        & 15                & 16500          & \nodata         & \nodata      & 6  \\
0908$+$422 & SDSS J091124.68$+$420255.9 & H        & 45                & 10250          & \nodata         & \nodata      & 6  \\
0911$+$059 & SDSS J091437.40$+$054453.3 & H        & 9.5               & 17000          & \nodata         & \nodata      & 6  \\
0912$+$536 & G 195$-$19                 & He       & $\sim 100$ & $7160\pm190$          & $0.75\pm0.02$   & 1.3301 d     & 31,4,32 \\
0922$+$014 & SDSS J092527.47$+$011328.7 & H        & 2.2               & 10000          & \nodata         & \nodata      & 1  \\
0930$+$010 & SDSS J093313.14$+$005135.4 & He (C$_2$H)? &  ?            & \nodata        & \nodata         & \nodata      & 1  \\
0931$+$105 & SDSS J093356.40$+$102215.7 & H        & 1.5               & 8500           & \nodata         & \nodata      & 6  \\
0931$+$507 & SDSS J093447.90$+$503312.2 & H        & 9.5               & 8900           & \nodata         & \nodata      & 6  \\
0941$+$458 & SDSS J094458.92$+$453901.2 & H        & 14                & 15500          & \nodata         & \nodata      & 6  \\
0945$+$246 & LB 11146                   & H        & 670               & $16000\pm2000$ & $0.90^{+0.10}_{-0.14}$ & \nodata & 33,34 \\
0952$+$094 & SDSS J095442.91$+$091354.4 & DQ       & ?                 & \nodata        & \nodata         & \nodata      & 6  \\
0957$+$022 & SDSS J100005.67$+$015859.2 & H        & 20                & 9000           & \nodata         & \nodata      & 1  \\
1001$+$058 & SDSS J100356.32$+$053825.6 & H        & 900               & 23000          & \nodata         & \nodata      & 6  \\
1004$+$128 & SDSS J100715.55$+$123709.5 & H        & 7                 & 18000          & \nodata         & \nodata      & 6  \\
1008$+$290 & LHS 2229                   & He (C$_2$H) & $\sim 100$     & 4600           & \nodata         & \nodata      & 35 \\
1012$+$093 & SDSS J101529.62$+$090703.8 & H        & 5                 & 7200           & \nodata         & \nodata      & 6  \\
1013$+$044 & SDSS J101618.37$+$040920.6 & H        & 7.5               & 10000          & \nodata         & \nodata      & 1  \\
1015$+$014 & PG 1015$+$014              & H        & $120\pm10$        & 14000          & $\log{g}=(8.0)$ & 98.74734 min & 36,37,38 \\
1017$+$367 & GD 116                     & H        & $65\pm5$          & 16000          & \nodata         & \nodata      & 39 \\
1026$+$117 & LHS 2273                   & H        & 18                & $7160\pm170$   & (0.59)          & \nodata      & 40 \\
1031$+$234 & PG 1031$+$234              & H        & $\sim 200-1000$   & $\sim 15000$   & \nodata         & 3.3997 hr    & 41,42,43 \\
1033$+$656 & SDSS J103655.38$+$652252.0 & DQ       & 4:                & \nodata        & \nodata         & \nodata      & 1  \\
1036$-$204 & LP 790$-$29                & He       & 50                & 7800           & $\log{g}=(8.0)$ & $24-28$ d    & 44,45 \\
1043$-$050 & HE 1043$-$0502             & He       & $\sim 820$        & $\sim 15000$   & \nodata         & \nodata      & 46,10 \\
1045$-$091 & HE 1045$-$0908             & H        & 16                & $10000\pm1000$ & $\log{g}=(8.0)$ & $2.7$ hr     & 47 \\
1050$+$598 & SDSS J105404.38$+$593333.3 & H        & 17                & 9500           & \nodata         & \nodata      & 1  \\
1053$+$656 & SDSS J105628.49$+$652313.5 & H        & 28                & 16500          & \nodata         & \nodata      & 1  \\
1105$-$048 & LTT 4099                   & H        & 0.0039            & $15280\pm20$   & $0.52\pm0.01$   & \nodata      & 17 \\
1107$+$602 & SDSS J111010.50$+$600141.4 & H        & 6.5               & 30000          & \nodata         & \nodata      & 1  \\
1111$+$020 & SDSS J111341.33$+$014641.7 & He ?     & ?                 & \nodata        & \nodata         & \nodata      & 1  \\
1115$+$101 & SDSS J111812.67$+$095241.4 & H        & 6                 & 10500          & \nodata         & \nodata      & 6  \\
1126$-$008 & SDSS J112852.88$-$010540.8 & H        & 3                 & 11000          & \nodata         & \nodata      & 1  \\
1126$+$499 & SDSS J112924.74$+$493931.9 & H        & 5                 & 10000          & \nodata         & \nodata      & 6  \\
1131$+$521 & SDSS J113357.66$+$515204.8 & H        & 7.5               & 22000          & \nodata         & \nodata      & 1  \\
1135$+$579 & SDSS J113756.50$+$574022.4 & H        & 9                 & 7800           & \nodata         & \nodata      & 6  \\
1136$-$015 & LBQS 1136$-$0132           & H        & $24\pm1$          & 10500          & \nodata         & \nodata      & 48,1 \\
1137$+$614 & SDSS J114006.37$+$611008.2 & H        & 58                & 13500          & \nodata         & \nodata      & 1  \\
1145$+$487 & SDSS J114829.00$+$482731.2 & H        & 33                & 27500          & \nodata         & \nodata      & 6  \\
1151$+$015 & SDSS J115418.14$+$011711.4 & H        & 32                & 27000:         & \nodata         & \nodata      & 1  \\
1156$+$619 & SDSS J115917.39$+$613914.3 & H        & 15.5              & 23000          & \nodata         & \nodata      & 1  \\
1159$+$619 & SDSS J120150.10$+$614257.0 & H        & 20                & 10500          & \nodata         & \nodata      & 6  \\
1203$+$085 & SDSS J120609.80$+$081323.7 & H        & 830:              & 13000          & \nodata         & \nodata      & 6  \\
1204$+$444 & SDSS J120728.96$+$440731.6 & H        & 2.5               & 16750          & \nodata         & \nodata      & 6  \\
1209$+$018 & SDSS J121209.31$+$013627.7 & H        & 13                & 10000          & \nodata         & \nodata      & 1  \\
1211$-$171 & HE 1211$-$1707             & He       & 50                & $\sim 12000$   & \nodata         & $\sim 2$ hr  & 10 \\
1212$-$022 & LHS 2534                   & He(DZ)   & 1.92           & 6000           & \nodata         & \nodata      & 49 \\
1214$-$001 & SDSS J121635.37$-$002656.2 & H        & 63                & 20000          & \nodata         & \nodata      & 13 \\
1219$+$005 & SDSS J122209.44$+$001534.0 & H        & 12                & 20000          & \nodata         & \nodata      & 13 \\
1220$+$484 & SDSS J122249.14$+$481133.1 & H        & 8                 &  9000          & \nodata         & \nodata      & 6  \\
1220$+$234 & PG 1220$+$234              & H        & 3                 & 26540          & 0.81            & \nodata      & 37 \\
1221$+$422 & SDSS J122401.48$+$415551.9 & H        & 23:               &  9500          & \nodata         & \nodata      & 6  \\
1231$+$130 & SDSS J123414.11$+$124829.6 & H        & 7                 &  8200          & \nodata         & \nodata      & 6  \\
1245$+$413 & SDSS J124806.38$+$410427.2 & H        & 8                 &  7000          & \nodata         & \nodata      & 6  \\
1246$-$022 & SDSS J124851.31$-$022924.7 & H        & 7                 & 13500          & \nodata         & \nodata      & 1  \\
1248$+$161 & SDSS J125044.42$+$154957.4 & H        & 20                & 10000          & \nodata         & \nodata      & 6  \\
1252$+$564 & SDSS J125416.01$+$561204.7 & H        & 52                & 13250          & \nodata         & \nodata      & 6  \\
1254$+$345 & HS 1254$+$3440             & H        & $9.5\pm0.5$       & $15000\pm4000$ & \nodata         & \nodata      & 50 \\
1309$+$853 & G 256$-$7                  & H        & $4.9\pm0.5$       & $\sim 56000$   & \nodata         & \nodata      & 26 \\
1312$+$098 & PG 1312$+$098              & H        & 10                & $\sim 20000$   & \nodata         & 5.42839 hr   & 38,25 \\
1317$+$135 & SDSS J132002.48$+$131901.6 & H        & 5                 & 14750          & \nodata         & \nodata      & 6  \\
1327$+$594 & SDSS J132858.20$+$590851.0 & H        & 18                & 25000          & \nodata         & \nodata      & 6  \\
1328$+$307 & G165-7                     & He (DZ)  & 0.65     & $6440\pm210$   & $0.57\pm0.17$   & \nodata      & 51 \\
1330$+$015 & G 62$-$46\tablenotemark{b} & H        & $7.36\pm0.11$     & 6040           & 0.25            & \nodata      & 52 \\
1331$+$005 & SDSS J133359.86$+$001654.8 & He (C$_2$H) ? &  ?           & \nodata        & \nodata         & \nodata      & 1  \\
1332$+$643 & SDSS J133340.34$+$640627.4 & H        & 13                & 13500          & \nodata         & \nodata      & 1  \\
1339$+$659 & SDSS J134043.10$+$654349.2 & H        & 3                 & 15000          & \nodata         & \nodata      & 1  \\
1349$+$545 & SBS 1349$+$5434            & H        & 760               & 11000          & \nodata         & \nodata      & 53 \\
1350$-$090 & LP 907$-$037               & H        & $\la 0.3$         & $9520\pm140$   & $0.83\pm0.03$   & \nodata      & 54,55 \\
1425$+$375 & SDSS J142703.40$+$372110.5 & H        & 30                & 19000          & \nodata         & \nodata      & 6  \\
1430$+$432 & SDSS J143218.26$+$430126.7 & H        & 30                & 24000          & \nodata         & \nodata      & 6  \\
1430$+$460 & SDSS J143235.46$+$454852.5 & H        & 30                & 16750          & \nodata         & \nodata      & 6  \\
1440$+$753 & EUVE~J1439$+$750\tablenotemark{b,e} & H & $14-16$           & $20000-50000$  & $0.88-1.19$     & \nodata      & 56 \\
1444$+$592 & SDSS J144614.00$+$590216.7 & H        & 7                 & 12500          & \nodata         & \nodata      & 1  \\
1452$+$435 & SDSS J145415.01$+$432149.5 & H        & 5                 & 11500          & \nodata         & \nodata      & 6  \\
1503$-$070 & GD 175\tablenotemark{b}    & H        & 2.3               &  6990          & $0.70\pm0.13$   & \nodata      & 4  \\
1506$+$399 & SDSS J150813.20$+$394504.9 & H        & 20                & 17000          & \nodata         & \nodata      & 6  \\
1509$+$425 & SDSS J151130.20$+$422023.0 & H        & 12                &  9750          & \nodata         & \nodata      & 6  \\
1516$+$612 & SDSS J151745.19$+$610543.6 & H        & 17                & 9500           & \nodata         & \nodata      & 1  \\
1531$-$022 & GD 185                     & H        & $0.035\pm0.016$\tablenotemark{f}   & $18620\pm285$  & $0.88\pm0.03$   & \nodata      & 55,57 \\
1533$+$423 & SDSS J153532.25$+$421305.6 & H        & 4.5               & 18500          & \nodata         & \nodata      & 1  \\
1533$-$057 & PG 1533$-$057              & H        & $31\pm3$          & $20000\pm1040$ & $0.94\pm0.18$   & \nodata      & 58,11,55 \\
1537$+$532 & SDSS J153829.29$+$530604.6 & H        & 12                & 13500          & \nodata         & \nodata      & 1  \\
1539$+$039 & SDSS J154213.48$+$034800.4 & H        & 8                 & 8500           & \nodata         & \nodata      & 1  \\
1603$+$492 & SDSS J160437.36$+$490809.2 & H        & 53                & 9000           & \nodata         & \nodata      & 1  \\
1639$+$537 & GD 356\tablenotemark{g}    & He       & 13                & $7510\pm210$   & $0.67\pm0.07$   & 0.0803 d     & 59,4,60 \\
1641$+$241 & SDSS J164357.02$+$240201.3 & H        & 4                 & 16500          & \nodata         & \nodata      & 6  \\
1645$+$372 & SDSS J164703.24$+$370910.3 & H        & 2:                & 16250          & \nodata         & \nodata      & 6  \\
1648$+$342 & SDSS J165029.91$+$341125.5 & H        & 3:                & 9750           & \nodata         & \nodata      & 6  \\
1650$+$355 & SDSS J165203.68$+$352815.8 & H        & 9.5               & 11500          & \nodata         & \nodata      & 1  \\
1658$+$440 & PG 1658$+$440              & H        & $2.3\pm0.2$       & $30510\pm200$  & $1.31\pm0.02$   & \nodata      & 61 \\
1702$+$322 & SDSS J170400.01$+$321328.7 & H        & 5                 & 23000          & \nodata         & \nodata      & 6  \\
1713$+$393 & NLTT 44447                 & H        & 1.3               & 6260           & \nodata         & \nodata      & 62 \\
1715$+$601 & SDSS J171556.29$+$600643.9 & H        & 4.5               & 13500          & \nodata         & \nodata      & 6  \\
1719$+$562 & SDSS J172045.37$+$561214.9 & H        & 21                & 22500          & \nodata         & \nodata      & 13 \\
1722$+$541 & SDSS J172329.14$+$540755.8 & H        & 33                & 16500          & \nodata         & \nodata      & 13 \\
1728$+$565 & SDSS J172932.48$+$563204.1 & H        & 28                & 10500          & \nodata         & \nodata      & 1 \\
1743$-$520 & BPM 25114                  & H        & 36                & $\sim 20000$   & $\log{g}=(8.0)$ & 2.84 d       & 63,64 \\
1748$+$708 & G 240$-$72                 & He       & $\ga 100$         & $5590\pm90$    & $0.81\pm0.01$   & $\ga 20$ yr  & 18,4,65 \\
1814$+$248 & G 183$-$35                 & H        & $\sim 14$         & $6500\pm500$   & $\log{g}=(8.0)$ & $\sim 50$ min $-$ few yr & 26,25 \\
1818$+$126 & G 141$-$2\tablenotemark{b} & H        & $\sim 3$          & $6340\pm130$   & $0.26\pm0.12$   & \nodata      & 66,40 \\
1829$+$547 & G 227$-$35                 & H        & $170-180$         & $6280\pm140$   & $0.90\pm0.07$   & \nodata      & 21,4 \\
1900$+$705 & Grw $+70^\circ 8247$       & H        & $320\pm20$        & 16000          & $0.95\pm0.02$   & \nodata      & 67,68,4 \\
1953$-$011 & G 92$-$40                  & H        & $0.1-0.5$         & $7920\pm200$   & $0.74\pm0.03$   & 1.44176 d    & 69,4,70 \\
2010$+$310 & GD 229                     & He       & $520$             & 18000          & $\log{g}=(8.0)$ & \nodata      & 46 \\
2043$-$073 & SDSS J204626.15$-$071037.0 & H        & 2                 & 8000           & \nodata         & \nodata      & 1  \\
2049$-$004 & SDSS J205233.52$-$001610.7 & H        & 13                & 19000          & \nodata         & \nodata      & 1  \\
2105$-$820 & L 24$-$52                  & H        & $0.043\pm0.01$ ?  & $10200\pm290$  & $0.75\pm0.03$   & \nodata      & 57,4 \\
2146$+$005 & SDSS J214900.87$+$004842.8 & H        & 10                & 11000          & \nodata         & \nodata      & 6  \\
2146$-$077 & SDSS J214930.74$-$072812.0 & H        & 42                & 22000          & \nodata         & \nodata      & 1  \\
2149$+$002 & SDSS J215135.00$+$003140.5 & H        & $\sim 300$        & 9000           & \nodata         & \nodata      & 1  \\
2149$+$126 & SDSS J215148.31$+$125525.5 & H        & 21                & 14000          & \nodata         & \nodata      & 1  \\
2215$-$002 & SDSS J221828.59$-$000012.2 & H        & 225               & 15500          & \nodata         & \nodata      & 1  \\
2245$+$146 & SDSS J224741.46$+$145638.8 & H        & 560               & 18000          & \nodata         & \nodata      & 1  \\
2316$+$123 & KUV 813$-$14               & H        & $45\pm5$          & $11000\pm1000$ & $\log{g}=(8.0)$ & 17.856 d     & 21,38 \\
2317$+$008 & SDSS J231951.73$+$010909.3 & H        & 1.5:              & 8300           & \nodata         & \nodata      & 6  \\
2320$+$003 & SDSS J232248.22$+$003900.9 & H        & 13                & 39000          & \nodata         & \nodata      & 13 \\
2321$-$010 & SDSS J232337.55$-$004628.2 & He       & 4.8               & 15000          & \nodata         & \nodata      & 1 \\
2329$+$267 & PG 2329$+$267              & H        & $2.31\pm0.59$     & $9400\pm240$   & $0.61\pm0.16$   & \nodata      & 71,4 \\
2343$+$386 & SDSS J234605.44$+$385337.7 & H        & 1000              & 26000          & \nodata         & \nodata      & 6  \\
2343$-$106 & SDSS J234623.69$-$102357.0 & H        & 2.5               & 8500           & \nodata         & \nodata      & 6  \\
2359$-$434 & LTT 9857                   & H        & 0.0031\tablenotemark{d} & $8570\pm50$ & $0.98\pm0.04$ & \nodata     & 17,72 \\
\enddata
\tablenotetext{a}{Polarization has been observed by \citet{sch2003}.}
\tablenotetext{b}{Unresolved double degenerate.}
\tablenotetext{c}{Polarization has been observed by \citet{sch2001b}.}
\tablenotetext{d}{The longitudinal field, $B_\ell$.}
\tablenotetext{e}{Also HS1440+7518, which was incorrectly labeled as HS1412+6115 in the original paper.}
\tablenotetext{f}{Magnetic field detection is based on the presence of a broadened H$\alpha$ core 
\citep{koe1998}, which can also be rotationally broadened, and therefore the magnetic
field of this star is uncertain.}
\tablenotetext{g}{H emission.}
\tablerefs{(1) \citet{sch2003}; (2) \citet{lie2003b}; (3) \citet{val2005}; (4) \citet{ber2001};
(5) \citet{ber1992}; (6) \citet{van2005}; (7) \citet{ach1992}; (8) \citet{lie1977}; (9) \citet{ach1991}; (10) \citet{sch2001b}; (11) \citet{ach1989};
(12) \citet{fri1997}; (13) \citet{gan2002}; (14) \citet{ven2003}; (15) \citet{fab2000};
(16) \citet{fin1997}; (17) \citet{azn2004}; (18) \citet{ang1978}; (19) \citet{duf2005}; (20) \citet{pra1989}
(21) \citet{put1995b}; (22) \citet{ven1999}; (23) \citet{sch1992};
(24) \citet{gio1998}; (25) \citet{put1997} (26) \citet{put1995};
(27) \citet{mar1984}; (28) \citet{fer1998};
(29) \citet{cla2001}; (30) \citet{wes2001}; (31) \citet{ang1977};
(32) \citet{ang1972}; (33) \citet{gle1994}; (34) \citet{lie1993}; (35) \citet{sch1999};
(36) \citet{wic1988}; (37) \citet{lie2003};
(38) \citet{sch1991}; (39) \citet{saf1989}; (40) \citet{ber1997};
(41) \citet{lat1987}; (42) \citet{ost1992};
(43) \citet{sch1986}; (44) \citet{bue1999}; (45) \citet{jor2002};
(46) \citet{wic2002}; (47) \citet{euc2005}; (48) \citet{fol1989}; (49) \citet{rei2001};
(50) \citet{hag1987}; (51) \citet{duf2006};
(52) \citet{ber1993}; (53) \citet{lie1994}; (54) \citet{sch1994}; (55) \citet{lie2005}; (56) \citet{ven1999b};
(57) \citet{koe1998}; (58) \citet{lie1985}; (59) \citet{fer1997b}; (60) \citet{bri2004}; (61) \citet{sch1992b};
(62) \citet{kaw2006}; (63) \citet{wic1979}; (64) \citet{weg1977}; (65) \citet{ber1999};
(66) \citet{gre1986}; (67) \citet{wic1988b};
(68) \citet{jor1992}; (69) \citet{max2000}; (70) \citet{bri2005};
(71) \citet{mor1998}; (72) This work }
\end{deluxetable}
\clearpage
\end{landscape}

\clearpage

\begin{deluxetable}{llrrrl}
\tabletypesize{\scriptsize}
\tablecaption{Stars removed from previous lists of magnetic white dwarfs.\label{table_non_mag_wd}}
\tablewidth{0pt} 
\tablehead{
\colhead{WD} & \colhead{Other Names} & \colhead{Spectral Type} & \colhead{$T_{\rm eff}$} & \colhead{M} & \colhead{References} \\
\colhead{}   & \colhead{}            & \colhead{}              & \colhead{(K)}           & \colhead{($M_\odot$)} & \colhead{} \\
}
\startdata
0000$-$345 & HE 0000$-$3430 & DC         & $6240\pm140$    & $0.77\pm0.11$ & 1,2 \\
0003$-$570 & HE 0003$-$5701 & DA $+$ dMe & $\ga 80000$ & \nodata           & 1 \\
0026$-$218 & HE 0026$-$2150 & DA $+$ dMe & \nodata         & \nodata       & 1 \\
0107$-$019 & HE 0107$-$0158 & DA $+$ dMe & \nodata         & \nodata       & 1 \\
0127$-$311 & HE 0127$-$3110 & DZ (?)     & \nodata         & \nodata       & 1 \\
0136$+$251 & PG 0136$+$251  & DA         & $39400\pm1200$  & $1.24\pm0.04$ & 3,4 \\
0338$-$388 & HE 0338$-$3853 & DA $+$ dMe & $87000-97000$   & \nodata       & 1 \\
2201$-$228 & HE 2201$-$2250 & DB         & \nodata         & \nodata       & 5 \\
\enddata
\tablerefs{(1) \citet{sch2001b}; (2) \citet{ber2001};
(3) \citet{sch1992b}; (4) \citet{ven1999}; (5) \citet{jor2001}}
\end{deluxetable}

\end{document}